\documentclass[structabstract]{aa}  
\usepackage{graphicx}
\usepackage{txfonts}
\usepackage{natbib}
\usepackage{booktabs}
\usepackage{upgreek}
\usepackage{setspace}
\usepackage{comment}
\usepackage{multirow}
\usepackage{float}
\usepackage{appendix}
\def \h2{\ifmmode{{\rm H}_2}\else{${\rm H}_2$}}

\def \nii {[N~{\sc ii}]6584~\AA}
\def \vhel{\ifmmode{~V_{{\rm HEL}}}\else{~$V_{{\rm HEL}}$}\fi}
\def \mycn18 {MyCn~18}

\begin{document}
   \title{VLT observations of the asymmetric Etched Hourglass Nebula, MyCn~18}

   \subtitle{}

   \author{N. Clyne
          \inst{1},
          M.P. Redman\inst{1}, M. Lloyd\inst{2}, M. Matsuura\inst{3}, N. Singh\inst{1}, J. Meaburn\inst{2}}

   \institute{Centre for Astronomy, School of Physics, National University of Ireland Galway, University Road, Galway, Ireland \\
              \email{n.clyne1@nuigalway.ie, matt.redman@nuigalway.ie}
	\and
{Jodrell Bank Centre for Astrophysics, School of Physics and Astronomy, University of Manchester, Oxford Road, Manchester M13 9PL, UK}
	\and
{Department of Physics and Astronomy, University College London, Gower Street, London WC1E 6BT, UK}      
	   }

   \date{\today}

 
  \abstract
   {The mechanisms that form extreme bipolar planetary nebulae remain unclear.}
   {The physical properties, structure, and dynamics of the bipolar planetary nebula, MyCn~18, are investigated in detail with the aim of understanding the shaping mechanism and evolutionary history of this object.}
   {VLT infrared images, VLT ISAAC infrared spectra, and long-slit optical Echelle spectra are used to investigate MyCn~18. Morpho-kinematic modelling was used to firmly constrain the structure and kinematics of the source. A timescale analysis was used to determine the kinematical age of the nebula and its main components.}
   {A spectroscopic study of MyCn~18's central and offset region reveals the detailed make-up of its nebular composition. Molecular hydrogen, atomic helium, and Bracket gamma emission are detected from the central regions of MyCn~18. ISAAC spectra from a slit position along the narrow waist of the nebula demonstrate that the ionised gas resides closer to the centre of the nebula than the molecular emission. A final reconstructed 3-D model of MyCn~18 was generated, which provides kinematical information on the expansion velocity of its nebular components by means of position-velocity (P-V) arrays. A kinematical age of the nebula and its components were obtained by the P-V arrays and timescale analysis.}
  {The structure and kinematics of MyCn~18 are better understood using an interactive 3-D modelling tool called {\sc shape}. A dimensional and timescale analysis of MyCn~18's major components provides a possible mechanism for the nebula's asymmetry. The putative central star is somewhat offset from the geometric centre of the nebula, which is thought to be the result of a binary system. We speculate that the engulfing and destruction of an exoplanet during the AGB phase may have been a key event in shaping MyCn~18 and generating of its hypersonic knotty outflow.}


\authorrunning{Clyne et al}
   \maketitle
%

\section{Introduction}
MyCn\,18\,(RA=13h39m35.07s,\,Dec=$-$67d22m51.7s\,(J2000); PN G307.5$-$04.9; IRAS\,13359$-$6707) 
is a beautiful example of an extremely tightly waisted bipolar
planetary nebula (PN). The famous HST images of \cite{Sahai1999}, who christened it
the Etched Hourglass Nebula, show the remarkable degree
of symmetry in the nebula and also the clear offset of the central
star along the nebular minor axis.  The question of how the diverse
range of PNe shapes are produced from what is presumably a
spherical AGB star \citep{soker1992,trammell1995,sahai1998,sahai2009} continues to intrigue astronomers and the extreme
case of MyCn~18 provides a stern test of any evolutionary model.

%
\begin{figure*}
\centering
\resizebox{\textwidth}{!}{
\includegraphics{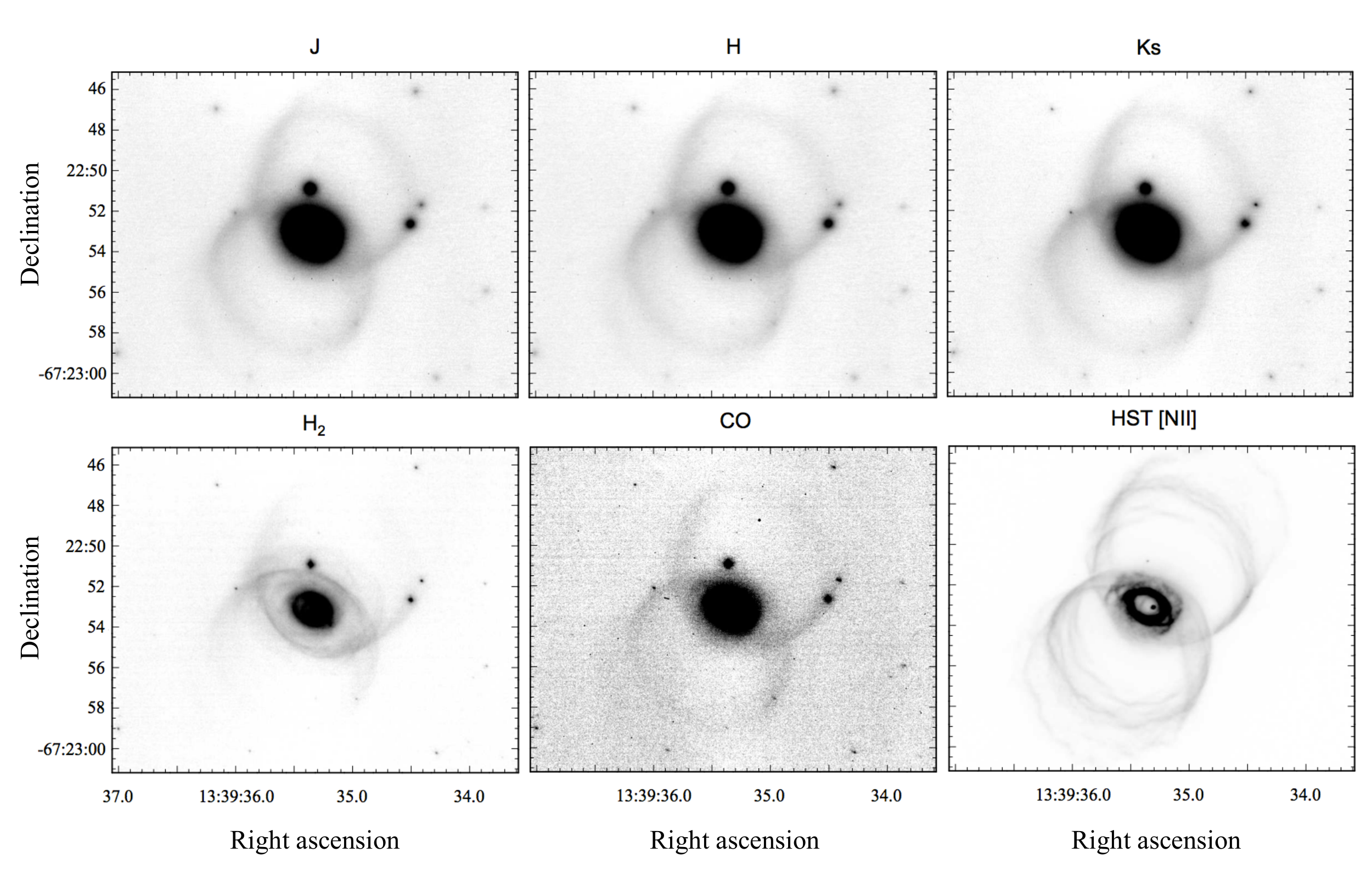}}
\caption{Images of the bulk of MyCn~18's nebulosity. Presented are a series of linear greyscale infra-red images obtained using the J, H, and Ks broadband filters, along with H$_{2}$ and CO images using the NB\_2.12 and IB\_2.30 filters, respectively. An archival HST \nii ~linear greyscale image is also included to compare and display the nebula's more distinct features.}
\label{fig:images}
\end{figure*}
%

The tight-waisted appearance of MyCn~18 led \cite{corradi1993} to compare 
it to two apparently similar symbiotic nebulae He2-104 and BI Cru.
However, they concluded that, other than the striking morphological similarity, there was no other
evidence pointing to MyCn~18 being a post-PN symbiotic
nebula and that it should tentatively be classed as a young PN, on the
basis of its position in optical and near-IR diagnostic diagrams.

\cite{Bryce1997} discovered hypersonic ($\sim$500~\rm km~s$^{-1}$) knots
situated on either side of the main hourglass nebula, approximately along 
the polar axis. It is interesting to note that these knots are
travelling much faster than the expansion velocity of the hourglass
shell ($\sim$50~\rm km~s$^{-1}$). Such speeds are similar to those found by
\cite{corradi1993} for the symbiotic nebulae He2-104 and BI Cru. 
\cite{o'connor2000} demonstrated that the knots of MyCn~18 had a
degree of point symmetry and that the radial speed of each knot
increased linearly with its distance from the central star.
They speculated that MyCn~18 might be morphologically related to the 
symbiotic nebulae.

The HST images of \cite{Sahai1999} revealed a
wealth of structural detail in the main nebula, which appears
to contain a second, inner, hourglass-shaped structure, as well as two central rings. Each of these structures 
has a slightly different geometric centre, none of which are coincident with the central
star. \cite{dayal2000} showed that the nebula is generally density-bounded
except in the region of the hourglass waist. They also inferred that
the etched arcs on the hourglass walls might indicate temporal
variations in the mass-loss rate of the progenitor star. \cite{Bains2002} 
found a similar geometric offset between the nebula and the central star 
when observing at radio wavelengths. Their spectral
index maps show that the nebular core appears to be more optically
thick than the lobe regions at lower frequencies (1384 and 4800 MHz).
The relatively low brightness temperatures were interpreted as a
result of beam dilution of a clumpy structure inferred a
filling factor of 0.15 at 1384~MHz. Their estimated total ionised mass
of the nebula is in the region 0.2--0.8 $\rm M_{\odot}$ at a distance of
2.4~kpc.

\cite{whitelock1985} presented JHK photometry for a sample of southern
PNe and showed that the major source of the near-IR
radiation from MyCn~18 is nebular in origin. Molecular hydrogen has been 
detected in many PNe \citep{hora1996,kastner1998,likkel2000,kastner2000,Davis2003,aleman2004,eyermann2004,matsuura2005,hora2005,wang2006,Matsuura2007,sellgren2008,ramos2008,tenenbaum2009}, and given that MyCn~18 is believed to be ionisation-bounded 
at its waist, it is likely that any available molecular hydrogen will be found around this region.
\cite{Sahai1999} pointed out that the expected K-band photospheric
flux (derived from optical photometry) is about half of the observed value,
indicating that the central star is probably surrounded by hot dust in
the waist region.

The hourglass shape of MyCn~18 is possibly formed by a fast, tenuous wind
expanding within a more slowly expanding cloud, which has a higher density near its 
equatorial region than near its polar regions. This snow-plow-like process was first introduced
by \citet{kwok1978}, as the interacting stellar winds (ISW) theory of nebular formation, and later as
the generalised interacting stellar winds (GISW) model \citep{kwok1982,balick1987}. 
The HST images of MyCn~18 reveal both intricate patterns (etchings) along its 
nebular walls and weak emission with filamentary components just beyond the walls.
MyCn~18 is a good example of an extreme bipolar PN with strong asymmetry or axi-symmetry.
This bipolar shape might also be the result of the formation of a dense equatorial accretion/excretion disk or of high-velocity collimated outflows, as discussed by \citet{Sahai1999}.

%
\begin{figure*}[ht]
\centering
\resizebox{\textwidth}{!}{
\includegraphics{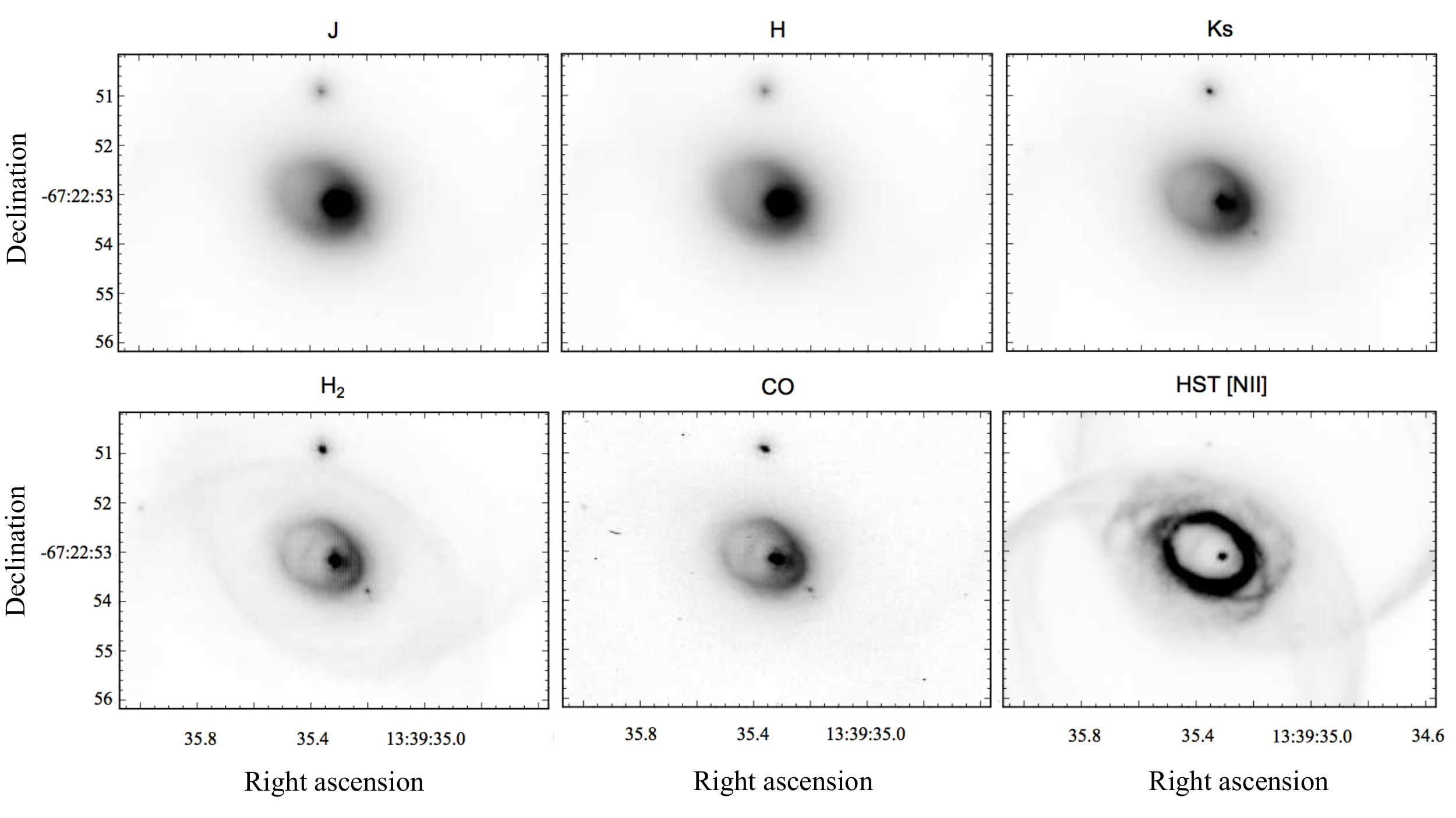}}
\caption{Linear stretches of the images in Fig.~\ref{fig:images}. The central region of each image is displayed at a higher contrast linear scale to show the features of interest not apparent in Fig.~\ref{fig:images}. The stretches, in comparison to those images in Fig.~\ref{fig:images}, are as follows: x10 for J, H, Ks and CO, x4 for H$_{2}$, and x2 for \nii.}
\label{fig:central}
\end{figure*}
%

The mechanism for generating extreme deviations from spherical
symmetry in planetary nebulae remains unsettled. Models and explanations
range from single stars with particular magnetic fields 
arrangements \citep{chevalier1994, garcia1999,pascoli2008,vlemmings2011,jordan2012,vlemmings2012}
to the view that almost all planetary nebulae are caused by binary
companions \citep{yungelson1993,soker2001,demarco2009,jones2010,douchin2012,jones2012}. 
The discovery that hot Jupiter exoplanets are common around stars means that large 
planetary mass companions can also be considered amongst the shaping mechanisms \citep{nordhaus2006}. Clearly it is of interest to search for a binary star in
MyCn~18. The identification of the central ionising star in MyCn~18
appears quite firm though it is clearly noticeable that the star is significantly
offset from the geometrical centre of the outer and inner hourglass, as identified by \citet{Sahai1999}.

In order to further our understanding of this striking nebula we have carried out an infra-red and spectroscopic study of MyCn~18 and used a 3-D morpho-kinematic code called {\sc shape}\footnote{{\sc shape} is accessible online at http://www.astrosen.unam.mx/shape/} \citep{steffen2011} to carefully reproduce the structural and spectral features of the nebula.

The paper is organised as follows: The observational data are presented in Sect.~\ref{observations}. Analyses of each topic are presented in Sect.~\ref{analysis}. This section includes: infra-red imaging, spectroscopic analysis, the modelling of optical kinematic data using {\sc shape}, a geometric dimensional analysis of the outer and inner regions of the nebula, and a timescale analysis of the nebula's three major components. A discussion and conclusions are found in Sect.~\ref{discussion} and Sect.~\ref{conclusions}, respectively.


\section{Observations and results}
\label{observations}


\subsection{Infra-red images}

A series of infra-red images of MyCn~18 were obtained in photometric conditions 
on 27 May 2003 using NAOS-CONICA 
\citep[NACO:][]{Rousset2003, Lenzen2003} 
on ESO's VLT(UT4) telescope.
NACO was used with the S27 camera (0.027\,arcsec per pixel and a field of 
view of $28\times28\rm~arcsec$).

Continuum images were obtained using the J, H, and Ks broadband filters. The
NB\_2.12 filter was used to image the 
\h2\ (1-0) S(1) emission line and the ${\rm IB}\_{2.30}$ filter was
used to look for evidence of CO 2.295~$\upmu$m emission.
The filters, exposure times, and image resolution are detailed in Table~\ref{table:obs}.

The central star of MyCn~18 [$m_{\rm v}$\,\,=\,\,14.9 and spectral type Of(C); as noted by \cite{Sahai1999} and \cite{Lee2007}, respectively]
was used as the AO reference star, with optical wavefront sensing via the 
VIS dichroic.  Optical seeing at the observatory was 0.4--1.1 arcsec FWHM 
during the observed period.
The observations were taken with the NACO\_img\_obs\_FixedSkyOffset
template using a 4 arcsec square jitter box and sky offset position
35 arcsec to the NE of the target position.

%

\begin{table}
\caption{\label{table:obs}Filters used for the NACO observations}
\centering
\begin{tabular}{lcccc}
\hline\hline\addlinespace[3pt]
Filter & $\lambda_{\rm c}~(\rm \upmu m)$ & $\Delta \lambda~(\rm \upmu m)$ & $t_{\rm exp}$
(\rm s) & \rm{Resolution} \\ \addlinespace[1pt]
\hline \addlinespace[3pt]
J & 1.265 & 0.25 & 1100 & $0.20\arcsec \times 0.19\arcsec$\\ \addlinespace[1pt]
H & 1.66 & 0.33 & 1100 & $0.18\arcsec \times 0.11\arcsec$\\ \addlinespace[1pt]
Ks & 2.18 & 0.35 & 1600 & $0.13\arcsec \times 0.11\arcsec $\\ \addlinespace[1pt]
${\rm NB}\_{2.122}$ &2.122& 0.022 & 3720 & $0.11\arcsec \times 0.09\arcsec$\\ \addlinespace[1pt]
${\rm IB}\_{2.30}$ & 2.30& 0.06 & 540 & $0.13\arcsec \times 0.09\arcsec$\\ \addlinespace[2pt]
\hline
\end{tabular} 
\tablefoot{A list of the filters used for the NACO observations, giving
the $\lambda_{\rm c}$ and $\Delta \lambda$ for each filter together with the
total on-target integration time and the resolution as measured from
the psf of the star to the northeast of the nebular core.}
\end{table}

%
                                     
The data were reduced using a tailored version of the ORAC-DR pipeline
\citep{Cavanagh2008} 
invoking the UKIRT imaging recipes
\citep{Currie2004}, which had been adapted to account for NACO's tiny point-spread function.
The final image from each filter was then aligned with the Ks image. These images are
shown in Fig.~\ref{fig:images}, along with the narrowband \nii\ HST image
\citep{Sahai1999}, which has been regridded to the NACO pixel scale and aligned with the Ks
image for consistency (The resolution of this image is about
$0.09\times0.09~\rm arcsec$).
Note that neither the \h2\ nor the CO images have
been continuum subtracted. The ${\rm NB}\_{2.122}$ filter, used for the \h2\
observations, produces a very irregular point spread function, and due
to this irregularity we were unable to create a satisfactory continuum
subtracted image. The stretch levels used for all images in both Fig.~\ref{fig:images} and 2 are tabulated in Table~\ref{table:stretch}.

%
\begin{figure}[ht]
\centering
\resizebox{\columnwidth}{!}{
\fbox{\includegraphics[width=0.5\columnwidth]{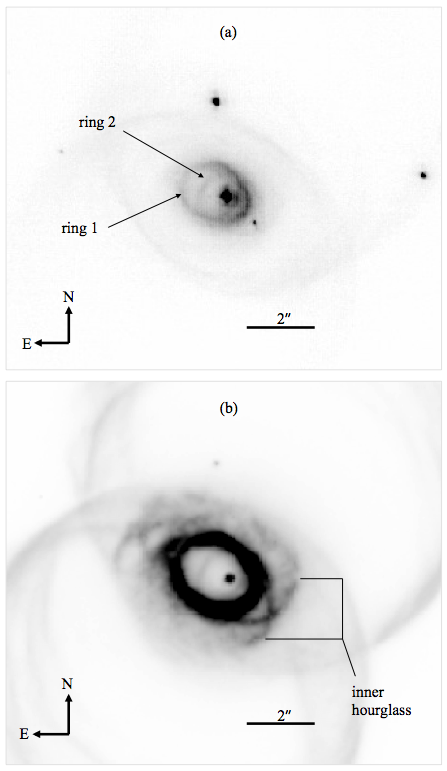}}}
\caption{Inner region of MyCn 18. (\textit{a}) Image of ring~1 and ring~2 in \rm H$_{2}$, and (\textit{b}) the inner hourglass in HST \nii\,.}
\label{fig:fig5}
\end{figure}
%

The images in both Figs. 1 \& 2 have been reduced in the standard way, except that no flux calibration has been applied. Each image represents a reduced mean frame uncorrected for exposure time.
The standard stars GSPC~P550--C (RA=10h33m51.9s,\,Dec=$+$04d49m05.0s) and GSPC S889-E (RA=22h02m05.73s,\,Dec=$-$01d06m01.0s) were observed with the J, H, and Ks filters but we were unable to derive an accurate flux calibration, however, the results were broadly compatible with the overall calibrations found in \citet{whitelock1985}.

The three continuum images in Fig.~\ref{fig:images} all show faint relatively smooth
emission from the edges of the main hourglass shell, together with a much
brighter central region. The \h2\ image in contrast shows a much
clearer thin, filamentary eye-shaped feature surrounding the bright inner region. 
Although the \h2\ image is not continuum subtracted, it is clear from the difference between
it and the corresponding broadband Ks image that the eye-shaped
feature is dominated by \h2\ line emission.

%

\begin{table}
\caption{\label{table:stretch}Linear greyscale cut off levels used in Figs. 1 and 2}
\centering
\begin{tabular}{ccccccc}
\hline\hline\addlinespace[3pt]
Fig & J & H & Ks & H$_{2}$ & CO & N~II \\ \addlinespace[1pt]
\hline \addlinespace[3pt]
1 & 0--20 & 0--20 & 0--16 & 0--15 & 0--10 & 0--800\\ \addlinespace[1pt]
\addlinespace[3pt]
2 & 0--200 & 0--200 & 0--160 & 0--60 & 0--100 & 0--1600 \\ \addlinespace[1pt]
\hline
\end{tabular} 
\tablefoot{Scaling was chosen to highlight the nebular features. The scales are in counts but are not flux calibrated. Individual frame exposure times were 10~s for J, H, and Ks, and 60~s for CO and H$_{2}$. The images shown in both figures are the combined and reduced images for each filter but with counts averaged to the exposure time for one image to give a better signal-to-noise.}
\end{table}

%

The bright central region of each image is reproduced in Fig.~\ref{fig:central} at
higher contrast to show the structure of the inner part of the nebula.
Note that Figs.~\ref{fig:images} and \ref{fig:central} are framed to progressively highlight the hourglass shape into the very central 
region of the eye.
The broadband images again show a relatively smooth structure,
dominated by an egg-shaped ring that is much brighter at its western,
pointed end, just southwest of the central star. This ring was identified in the HST WFPC2 narrow-band
optical images of \cite{Sahai1999}
as ring 1. The central star is offset towards the narrow end of the
ring, as noted by \cite{Sahai1999}. 
An inner loop of emission corresponding to ring 2
is also apparent, particularly in the \h2\ image.

Fig.~\ref{fig:fig5} shows the central region of MyCn~18 in (a) \h2,
identifying ring~1 and ring~2, and (b) in \nii, revealing the nebula's inner hourglass. Ring 1 is defined as the waist of the outer hourglass and ring 2 appears to be part of the inner hourglass, which is a higher excitation ring due to its close proximity with the central star \citep{Sahai1999}. 

%
\begin{figure}
\centering
\resizebox{\columnwidth}{!}{
\fbox{\includegraphics[viewport=23 252 573 620,clip]{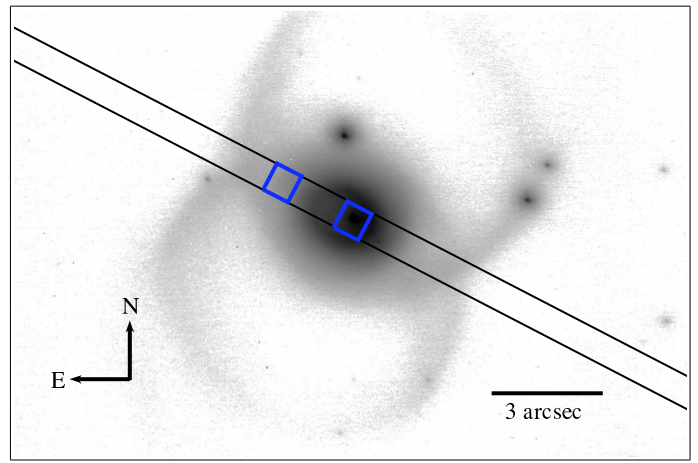}}}
\caption{Approximate ISAAC slit position displayed on NACO  Ks image.
Two boxes show the positions whose spectra have been displayed in 
Fig.~\ref{fig:NIRspec}.}
\label{fig:isaac-slit}
\end{figure}
%

%
\begin{figure*}
\centering
\includegraphics[width=17.5cm]{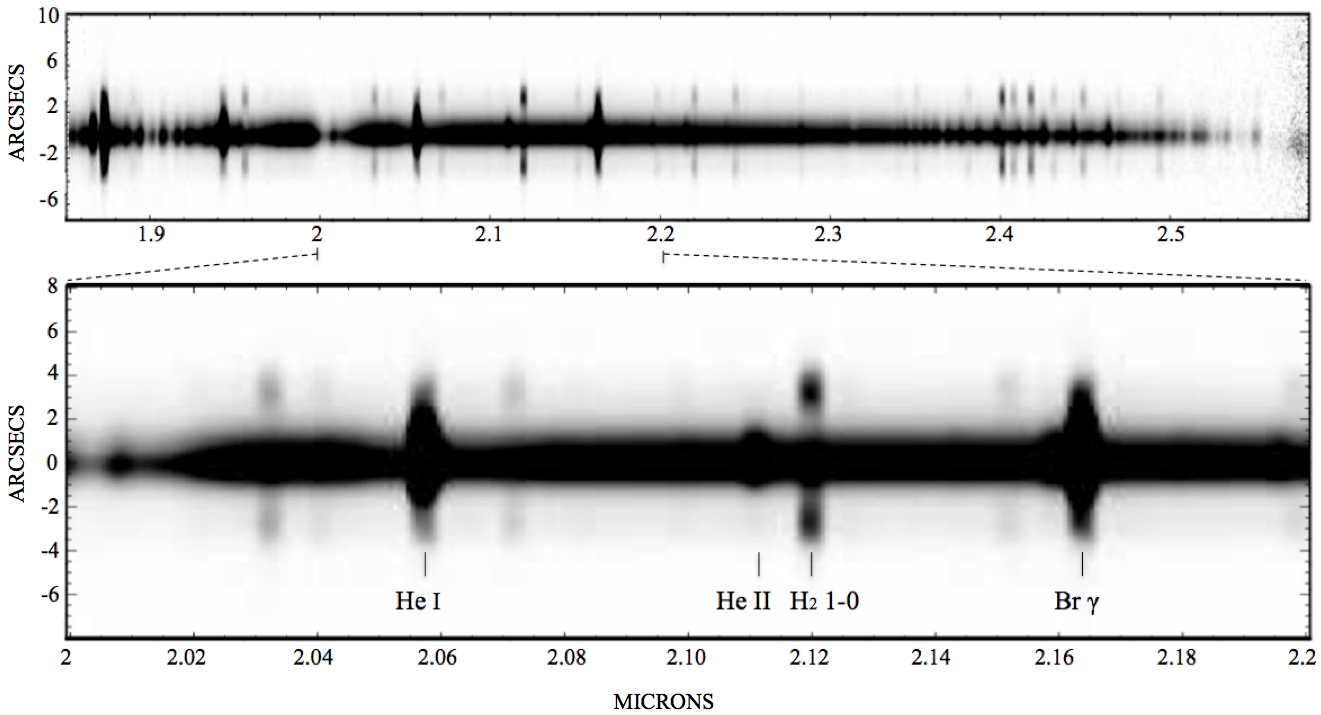}
\caption{A spatially resolved VLT ISAAC (outflow) spectrum. The spectrum, taken across the minor axis of the main hourglass shell, is shown at high contrast with the most prominent atomic and molecular emission lines visible. Two-micron spectrum (\textit{top}), 2.25 arcsec from the central region along the nebula's minor axis, shows a plethora of strong He recombination and H$_{2}$ vibrational lines. Below is an enlarged view of the region of this spectrum over the range 2--2.2 micron. More details of this spectrum are shown in Fig.~\ref{fig:NIRspec} }
\label{fig:isaac}
\end{figure*}
%

%

\begin{figure*}
\centering
\includegraphics[width=16.4cm]{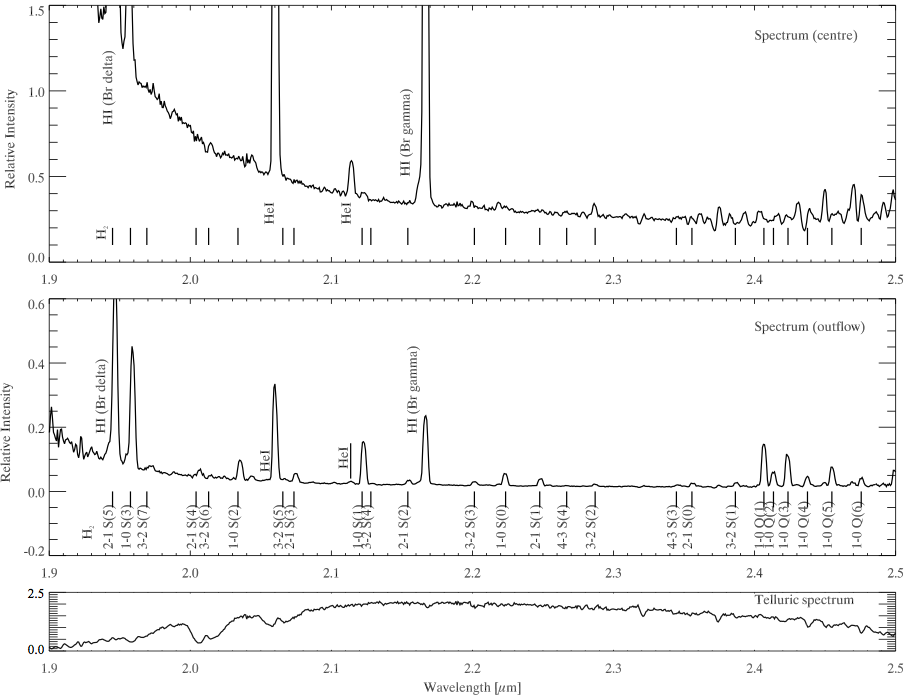}
\caption{Two-micron spectra of MyCn~18. Spectra for the central (\textit{top}) 0.75 and offset (\textit{middle}) 2.25 arcsec region are illustrated. Major H\,\small{I} \normalsize and He\,\small{II} \normalsize lines are indicated, and the expected positions of H$_2$  lines are marked. The bottom panel shows the atmospheric features.}
\label{fig:NIRspec}
\end{figure*}

%

%
%

\subsection{Infra-red spectroscopy}

Two-micron spectra of MyCn~18 were obtained
with ISAAC \citep{Moorwood1999} 
at the ESO Very Large Telescope (VLT) on 29 April 2004.
An ISAAC short wavelength band was equipped with the Rockwell 
$1024\times1024$ array with pixel scale 0.145 arcsec in imaging 
mode and 0.147 arcsec in spectroscopic mode. 
The weather was clear, optical seeing varied between 0.92--1.37 arcsec 
and seeing at the K-band was approximately 0.75 arcsec during the 
acquisition of the target.
The slit was placed along the minor axis of the nebula at a PA of
62.45$^{\circ}$ to the east (see Fig.~\ref{fig:isaac-slit})
and the spatial scale was 0.15 arcsec per pixel.
Low spectral resolution (LR) mode was used, and the slit width was 0.8 arcsec, 
resulting in a wavelength resolution ($R= \lambda / \Delta \lambda$) of 750. 
The telescope was nodded by 60 arcsec along the slit to cancel the sky background.
Jitter was used along the slit to
minimise the influence of hot pixels.
An exposure of twilight sky was used for flat-fielding.  
The exposure time was 120\,s~$\times$~16 on source.

The approximate ISAAC slit location is shown in Fig.~\ref{fig:isaac-slit} and the orientations 
of the slit are from northeast to southwest, which follow from top to bottom in the figure. 
The two boxes across the ISAAC slit represent 
the regions of the centre and outflow spectrum.
The data were reduced using {\sc idl} and the {\sc eclipse} package.
The telluric standard was the G2V star Hip\,053690; \citet{Pickles1998} calculated the spectrum of this star.
The wavelength calibration is based on exposures of an Ar+Xe arc lamp
with the same wavelength setting as the target observations.
Additionally, OH skylines 
\citep{oliva1992,maihara1993,Rousselot2000} 
and telluric absorption lines \citep{hadrava2006} were used for the fine adjustment of 
the wavelength calibration.

The 2-D spectrum, shown at high contrast in
Fig.~\ref{fig:isaac}, has been binned 
by 5 pixels along the slit length and indicates the most prominent atomic and molecular
emission lines. The spectrum is seen to contain
both continuum emission from the central region of MyCn~18 and line
emission from the surrounding nebular shell. A spatially extended nebular continuum 
is clearly shown to be about 2 arcsec in extent and slightly offset with respect to the nebula's geometric centre.
The continuum emission has a faint extension to the east of
the array. Two types of emission lines can be identified: one which is
brightest in the core region and the other which is relatively faint
in the core region but is brighter at each end, where the slit crosses
the eastern and western edges of the nebula. This spectrum shows the 
spatial extent of \h2, He, and Br $\gamma$ lines but is not corrected 
with telluric standard.

In Fig.~\ref{fig:NIRspec}, a K-band spectrum of MyCn~18 extracted
from the central region is compared to a spectrum extracted from the
nebular outflow region. The identifications of representative lines are 
also marked in the figure.
In the central region, continuum is dominant, and both the central 
star and free-free and bound-free
contribute to this continuum, as found in the images. This continuum contributes to approximately 8\%
of the total in-band flux in the Ks-band. 
There are strong H (Br~$\gamma$ and Br~$\delta$) and He 
recombination lines.  Two H$_2$ lines can be identified at  
1.95 $\upmu$m 1--0 S(3) and 2.12 $\upmu$m 1--0 S(1).
The identifications of the lines beyond 2.35 $\upmu$m are uncertain, but some H$_2$ lines 
are present.
In contrast, a forest of H$_2$ lines is found in the outflow region, in addition
to the H and He recombination lines. 

The spectra shown in Fig.~\ref{fig:isaac} and \ref{fig:NIRspec}
demonstrate the spatial variations of the spectra across the object.
The spectrum at the central region shows a strong underlying 
continuum across K-band spectra, particularly below the 2.2 $\upmu$m 
wavelength. Strong H and He recombination lines are present in this region 
but \h2 lines are undetected. In contrast, the spectrum at the offset 
position, near the inner hourglass, is line dominated with some continuum. 
Here, He recombination lines are as strong 
as H recombination lines, but notably, much stronger \h2 lines are 
detected across this region. These spectra reveal the dominant 
contributors to the \h2 image (from Fig.~\ref{fig:images}) taken by the NACO. 
In the central region, there is a bright core found in the NB\_2.122 image 
but this is due to continuum rather than \h2. This image 
shows clear bipolar shaped outflows, which outline the hourglass shape 
and this part is likely to be formed by \h2 emission lines. CO has a band of lines 
near/at the 2.3 $\upmu$m line, which correspond to a large number of P- and R-branch 
vibrational-rotational transitions.

%
%

\subsection{Optical Images \& Spectroscopy}

To aid our analysis of MyCn~18 we obtained optical images from the Hubble Legacy Archive\footnote{Based on observations made with the NASA/ESA Hubble Space Telescope, obtained from the data archive at the Space Telescope Institute. STScI is operated by the association of Universities for Research in Astronomy, Inc. under the NASA contract NAS 5-26555.} and ESO Archive, as well as MES and EMMI long-slit spectroscopic data.

Position-velocity (P-V) arrays of \nii~line profiles from long-slit spectra were obtained along the major and minor axis of MyCn 18. These long-slit spectra were not flux calibrated and they have not been relatively calibrated either because we are primarily interested in the kinematics. Both sets of spectra are displayed with a linear scale. 

The major nebula axis long-slit spectra were taken with the Anglo-Australian telescope (AAT) using the Manchester Echelle Spectrometer \citep[MES;][]{meaburn1984} between 26 and 28 March 1999. A slit width of 150~$\upmu\rm m$, which is equivalent to 1 arcsec, was used and limits the resolution in the spectral direction to be 11~km~s$^{-1}$. In the spatial direction, seeing was about 1 arcsec.
The minor nebula axis long-slit spectra were obtained using ESO's Multi-Mode Instrument \citep[EMMI;][]{dekker1986}, which was attached to the ESO New Technology telescope (NTT), and spectra were taken on 23 March 1994. The seeing conditions are the same for those of the AAT, described above. For additional details on the MES and EMMI spectra, see \citet{o'connor2000} and \citet{corradi1993}, respectively.

%
%

\section{Analysis}
\label{analysis}
\subsection{Infra-red images}

The images were analysed, using {\sc iraf}\footnote{IRAF is distributed by the National Optical Astronomy Observatories, which are operated by the Association of Universities for Research in Astronomy, Inc., under cooperative agreement with the National Science Foundation.}, for possible companion stars to the nebula's primary.
There is significant extinction and diffuse continuum emission to the centre of MyCn~18 \citep{Bryce1997}, which makes it difficult to resolve any close binary companions to the primary star. We believe the star located just SW of the central star in Fig.~\ref{fig:fig5}a has not been observed before. However, this star appears unlikely to be a candidate for association with the primary of MyCn~18 since it lies $\sim$0.9 arcsec ($\sim$2,900 AU for a 3.2~kpc nebula distance) from the central star. Given that there are two other stars in the field of Fig.~\ref{fig:fig5} then this newly observed star is also probably a field star; even if it were a companion, at this distance from the central star, it would have no direct dynamical effect on the shaping of the nebula.

Infrared contouring was later performed in and around the central region to search for even closer binary candidates. The astronomical data
processing software {\sc starlink} \citep{disney1982} and its {\sc kappa} \citep{currie2000} package were used to carry 
out the contouring. The reduced images in the H and Ks bands, as well as that in \h2, revealed some non-circular extension from the central star (see Fig.~\ref{fig:fig5}a). At first we thought this might be the result of a very close binary star, however, this non-point source appearance could be due to either (i) the aligning
and co-adding of the images, (ii) poor psf as a result of adaptive optics, (iii) knotty emission from a recent ejection event or (iv) a line of sight star.
Due to the large intensity of continuum surrounding the central star, it is not presently possible to distinguish further between these possibilities. Further high resolution observations of the central region would be required to reveal the nature of this source.

%

\begin{figure}
\centering
\resizebox{\columnwidth}{!}{
\includegraphics*[angle=90,viewport=59 63 524 679,clip]{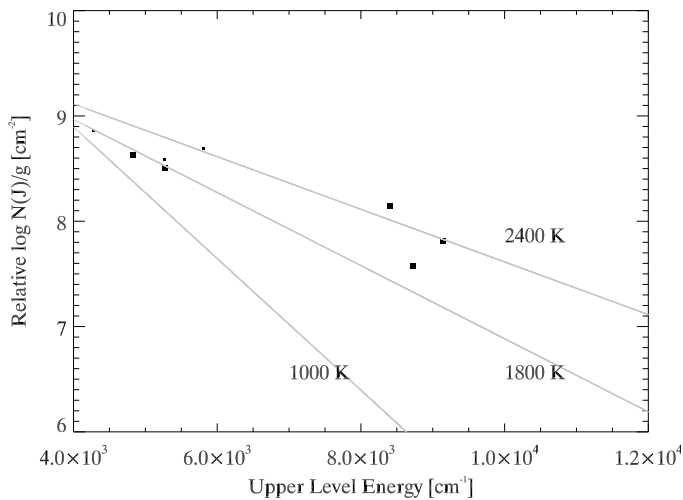}}
\caption{\normalsize An energy diagram of H$_2$ at 2.25 arcsec to the northeast 
of the central region. Squares show the measured line intensities and the lines show
expected H$_2$ line intensities for a given temperature.
The slopes of the lines indicate the temperature of the H$_{2}$ and shows the 
modelled energy distributions for 1000 K, 1800 K, and 2400 K,
where the vertical scales are arbitrary.
The H$_2$ excitation temperature is above 1800 K.}
\label{fig:h2-energy}
\end{figure}

%

\subsection{Spectroscopic analysis}
\label{spec} 

There are prominent atomic recombination lines from the central region extending outwards beyond the main continuum band. The H$_{2}$ lines are clearly different in spatial extent and peak at about $\pm3$ arcsec from the star. Comparing this spatial cut to the images, it can be seen that the H$_{2}$ emission comes from the cusps of the lenticular H$_{2}$ feature, which is most likely the line-of-sight overlap of the two lobes of the hourglass. \citet{Sahai1999} noted that the expected K-band flux (from the star) is about half the level observed by \citet{whitelock1985}  and suggested this may indicate the presence of hot dust in the equatorial region. Our spectra show that this excess emission can be explained instead by the strong free-free and atomic line emission from the compact ionised zone \citep{dayal1997,koller2000,volk2003,ramos2012}. 
Silicates typically have a dust condensation temperature of $\sim$1000~K, therefore, the peak of the dust emission would appear at a longer wavelength (2.9~$\upmu\rm m$) than what is observed in the spectra, which is actually $< 1.9~\upmu\rm m$. According to \cite{corradi1993}, MyCn 18 peaks at about 60 $\upmu$m ($\sim$48 K) and shows a typical double hump from its spectral energy distribution, indicating the presence of cool dust. Also, the light scattered by dust makes for an insignificant contribution to the nebula's central emission as noted by \cite{Sahai1999}.

%
\begin{figure}
\centering
\resizebox{\columnwidth}{!}{
\includegraphics[width=\columnwidth]{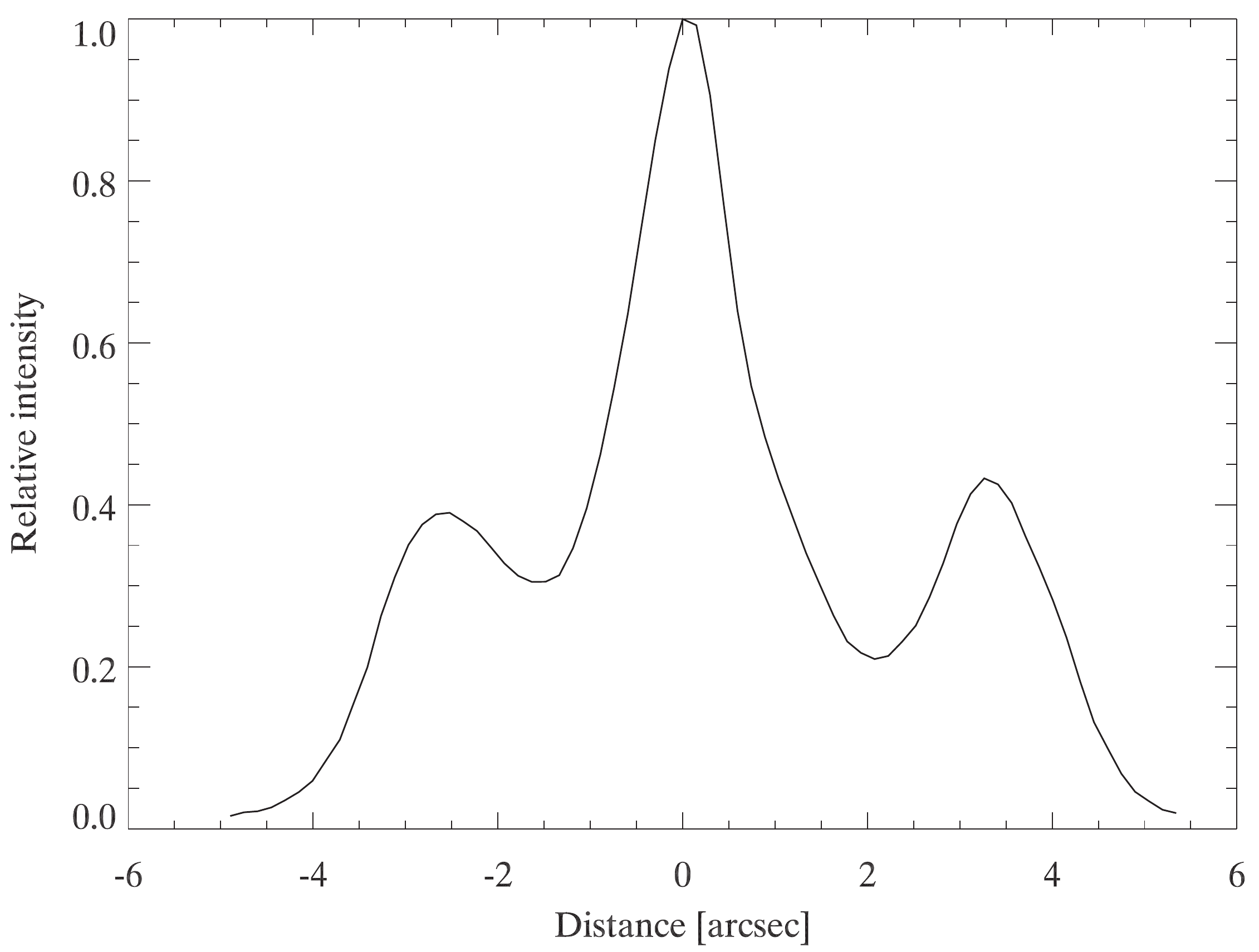}}
\caption{A spatial cut through the H$_{2}$ 1-0 S(1) line in Fig.~\ref{fig:isaac} along the 2.122 micron pixel. The intensity distribution is shown and scaled at the peak.}
\label{fig:cut}
\end{figure}
%

The resulting spectrum is the product of the nebula intensity gradient with a stratified ionisation structure, in which the least
ionised layers are found furthest from the central star.
The infra-red spectroscopy shows clear ionisation stratification in that the He, H, and H$_{2}$ lines are found progressively further from the central star into the putative surrounding molecular torus. The nebula is likely to be ionisation bounded in the equatorial direction but since the high-speed knots are ionised the nebula is more likely to be density bounded in the polar direction \citep{dayal2000,Bains2002}.

%
\begin{figure*}
\begin{minipage}{0.99\linewidth}
\centering
\includegraphics[width=15.2cm]{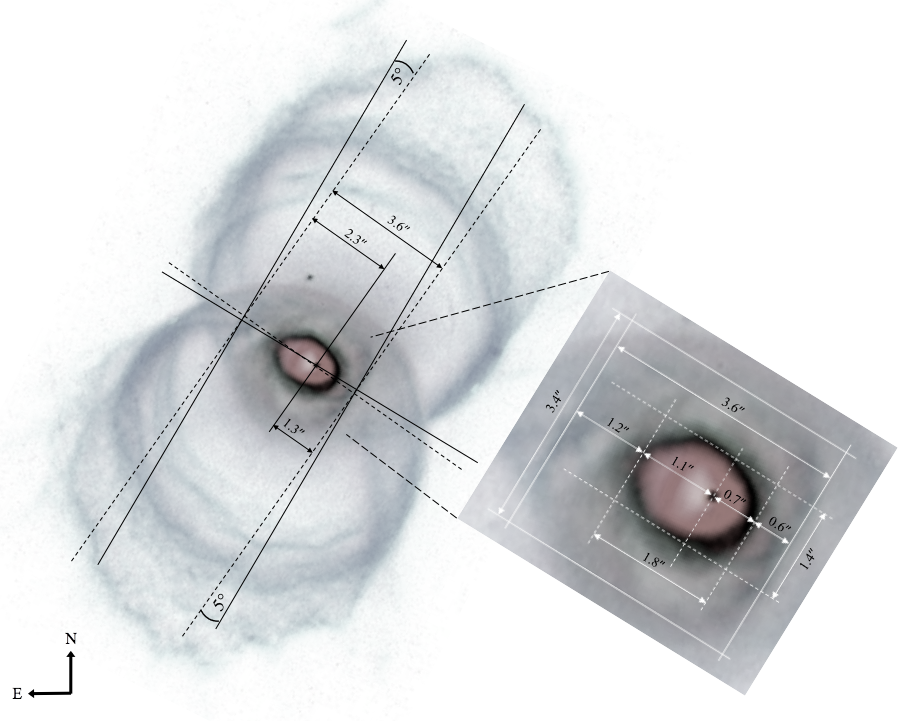}
\caption{An HST \nii~image of MyCn~18. Shown is a geometric dimensional analysis of its inner and outer regions. The asymmetry of the nebula is obvious from the figure, with the inner hourglass being offset and slightly tilted ($\approx$5$^{\circ}$) with respect to the main hourglass shell and the central star is offset from the nebula's geometric centre.\newline}
\label{fig:geo}
\end{minipage}
\end{figure*}
%

%
\begin{figure*}
\begin{minipage}{0.99\linewidth}
\centering
\includegraphics[width=18.3cm]{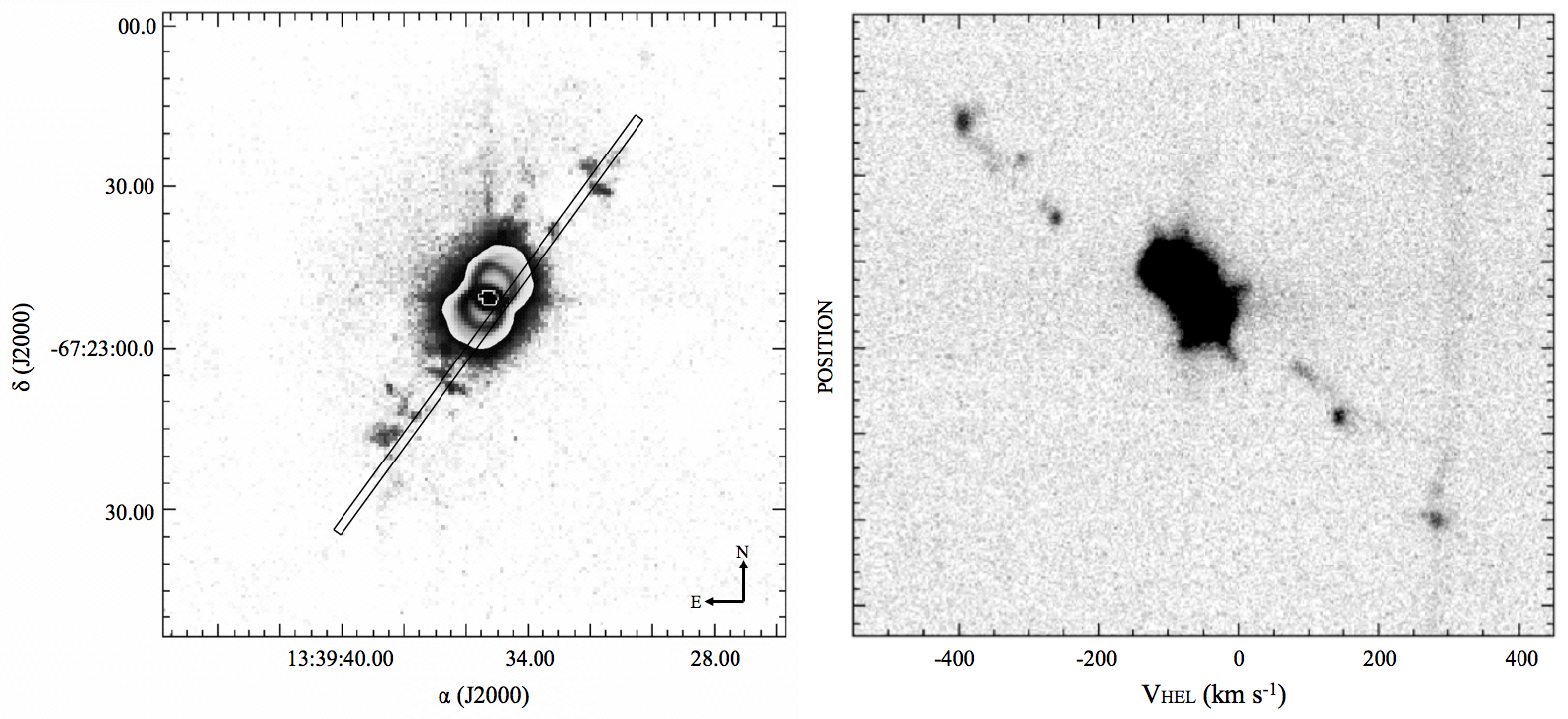}
\caption{Images of the hypersonic knotty outflow of MyCn 18. \textit{Left}: a continuum subtracted image of MyCn~18 revealing the relative positions of the high-speed knots. \textit{Right}: a P-V array for a long-slit placed down the nebula's major axis and through various knots, located north and south of the nebular lobes. The slit position shown on the left is a rough estimate of that used to produce the P-V array on the right. The knots south of the nebula (and through the slit) in the continuum subtracted image present the knots in the top left of the P-V array, and conversely, the knots north of the nebula present those in the bottom right. Both images in the figure are adopted from \citet{o'connor2000}.}
\label{fig:knots}
\end{minipage}
\end{figure*}
%

Fig.~\ref{fig:h2-energy} shows the energy diagram of H$_2$ lines in 
the outflow. The slopes, derived from the H$_{2}$ lines, constrain the range of possible temperatures of the H$_{2}$ emitting gas under the LTE assumption. This gives a temperature range of 1800--2400~K, though higher excitation lines may deviate from LTE \citep{Matsuura2007}.
The H$_2$ excitation temperature, estimated from 
rotation-vibrational lines, is as high as those estimated from similar 
lines of the Helix Nebula \citep[1800 K; ][]{Matsuura2007} 
and the centre of NGC 6302 \citep[1845 K;\,][]{Davis2003} 
but much higher than temperatures estimated from rotation lines
\citep[below 1000 K;\,][]{Cox1998, Bernard-Salas2005}.
It appears that rotation-vibrational temperatures are always 
higher near the central region of PNe.
This might require an extra heating mechanism such as UV 
excitation \citep{Henney2007} or shock excitation of H$_2$ \citep{Burton1992}.

The Ks band image in Fig.~\ref{fig:central}, which is dominated by the nebular continuum, shows an extent of about 2 arcsec across the central region. The spatial extent of the nebular continuum can also be readily seen in Fig.~\ref{fig:isaac}. Visually, it can be seen that the H$_{2}$ 1-0 line emission (at 2.12 $\upmu\rm m$) is found closer to the continuum in the negative spatial (SW) direction than the positive spatial (NE) direction, as shown in Fig.~\ref{fig:cut}. This corresponds with the asymmetry seen in the continuum images of Fig.~\ref{fig:central}.

\subsection{Geometric and timescale analysis of MyCn~18's nebular structures}
\label{dimensional} 

The nebula has noticeable deviations from axial symmetry as illustrated in Fig.~\ref{fig:geo}.
The figure shows the relative offsets of the various features within the nebula. As shown, the central star is 
not centred on the nebula's major axis and is clearly offset along the minor axis. The star is also offset from the centre of ring 1 and is closer to the centre of ring 2 (see Fig.~\ref{fig:fig5}a).
A close observation of the inner hourglass shows a slight tilt of its structure with respect 
to the outer hourglass. 
The image on the right of Fig.~\ref{fig:geo} shows 
the relative offsets of the inner hourglass, the nebular waist (ring 1), and the central star with respect to the nebula's geometric centre. In the case of the central star, it remains unclear as to why it is offset to the west of the nebular waist. An explanation of this offset may involve a binary companion.

In relation to MyCn~18's hypersonic knotty outflow, shown in Fig.~\ref{fig:knots}, the paths taken by the knots are not symmetric 
about the axis of symmetry of the nebula's main hourglass \citep{o'connor2000}. A closer inspection illustrates the scattering of the knots, with the northern cluster of knots found predominantly to the right of the main hourglass' axis of symmetry, and conversely, the southern cluster positioned to the left of this axis. The knots appear better aligned with the inner hourglass' axis of symmetry, thus supporting the hypothesis of a connection between these components.

%
%

The morphology and kinematics of MyCn~18 were better understood by applying timescales 
to each of its major components. These components include the main hourglass 
structure, the inner hourglass, and the high-speed interstellar knots. A distance of 
$2.4~{\rm kpc}$ is routinely quoted for MyCn~18, which is inferred from a
statistical method by Shklovsky \citep{schwarz1992}. A more recent re-analysis of this method 
by \citet{stanghellini2008} finds in particular that values derived in this way, by \cite{corradi1993}, underestimate the distance for butterfly type planetary nebulae. 
They find a value for the distance to be 3.2~kpc. Other distances quoted in the
literature range from 1~kpc to~3~kpc. 

The fastest knots are travelling with speeds of $\sim$500~km~s$^{-1}$ with the kinematical age of the fastest knot being $\sim$1250 yrs \citep{Bryce1997,o'connor2000}. \citet{dayal2000} presented 
details of the velocities and corresponding timescales along different latitudes of the 
main hourglass and found the kinematical age of the nebula to be 
$\sim$2500 yrs. 

While the radius, $R$, of the nebula will vary linearly using the adopted distance, the expansion velocity follows a weaker power law in $R$ \citep[$V_{\rm exp} \propto R^{(0.6\pm0.4)}$:][]{dayal2000}, hence the nebula expansion timescale is less sensitive than the estimated timescale of the ballistic ejection of the hypersonic knots.
Our value for the kinematical age of the fastest knot, using the 3.2~kpc distance, is $\sim$1650 yrs. With regards to the kinematical age of the nebula, the value obtained was $\sim$2700 yrs.
 
The shape and structure of the inner hourglass appears similar to that of the main hourglass, which
suggests that the mechanism 
that gave rise to the main hourglass might also be the same for that of the inner hourglass. 
Assuming this theory, the inner hourglass should follow a
similar velocity law ($V_{\rm exp} \propto R^{0.6\pm0.4}$) to that of the main hourglass. In this case, the expansion velocity of the 
inner hourglass would increase with increasing latitude along its walls, and is estimated to 
have an expansion velocity of 30 $\pm$ 5~km~s$^{-1}$ according to the synthetic P-V arrays generated from our {\sc shape} model. This corresponds to a timescale of 1750 $\pm$ 250~yrs. The interstellar knots have a similar kinematic age, thus implying that there might be a connection between them and the nebula's inner hourglass.

%
\begin{figure*}
\centering
\includegraphics[width=17.4cm]{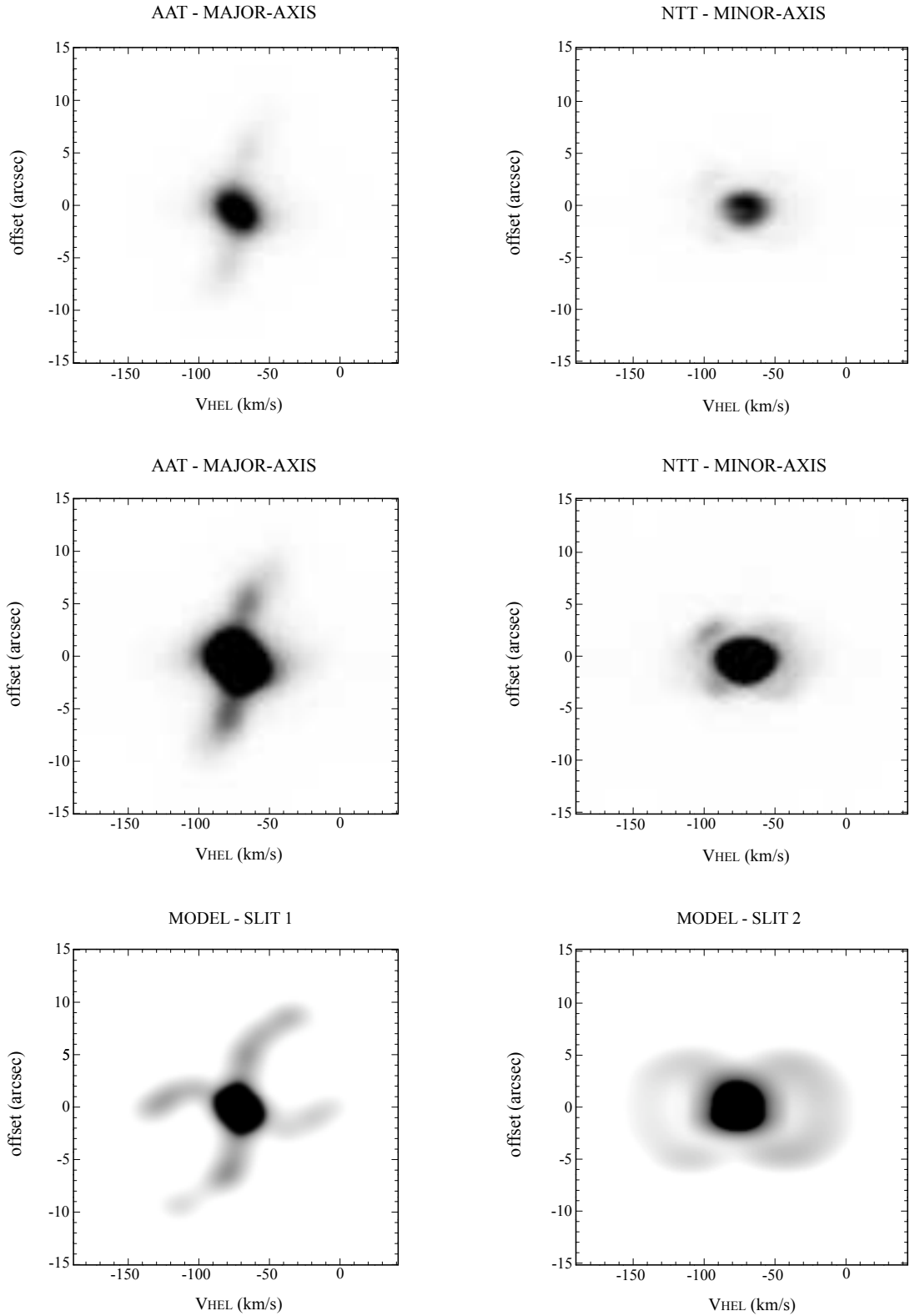}
\caption{MES, EMMI, and synthetic P-V arrays. The arrays on the left column (labelled AAT) are those for a long-slit applied to the major axis of the nebula using MES, whereas the right column (NTT) shows the arrays for a long-slit placed across the minor axis using EMMI (see Fig.~\ref{fig:shape}b for these slit positions). The P-V arrays on the top panel and those on the central panel are the same, except using different stretches of the intensity scale. Shown at the bottom (MODEL) panel are the corresponding synthetic P-V arrays, which were generated using the {\sc shape} model.}
\label{fig:mes}
\end{figure*}
%

%
\begin{figure*}
\centering
\includegraphics[width=14.2cm]{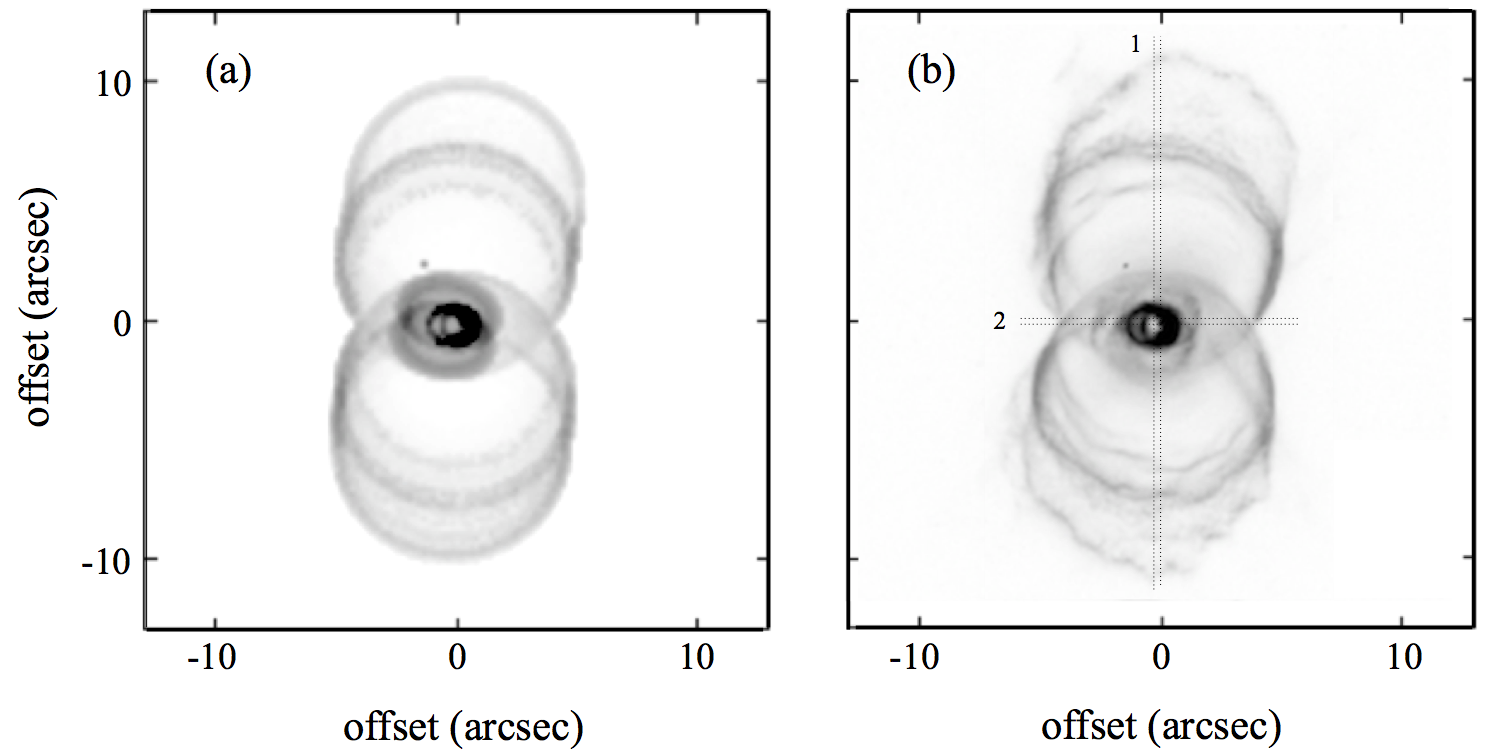}
\caption{The resultant {\sc shape} model of MyCn~18's nebular morphology and an HST\nii~image scaled for comparison. Also shown are the long-slit positions used for the observed and synthetic P-V arrays in Fig.~\ref{fig:mes}, labelled 1 (slit 1 down the major-axis) and 2 (slit 2 across the minor-axis). Both slit widths have a value of 0.8 arcsec.}
\label{fig:shape}
\end{figure*}

%

\subsection{Morpho-kinematic modelling of MyCn~18: {\sc shape} analysis}
\label{modeling}

While the hourglass morphology appears obvious, the
significant inclination of MyCn~18 means that the structure and
dynamics of the crucial region around the central star is not clear. Following the treatment of Hubble 12 in Vaytet et al (2009), the astrophysical modelling program {\sc shape} (version 4.51) was used to create the morpho-kinematics of MyCn~18 that were characterised by \citet{Bryce1997, Sahai1999, dayal2000, o'connor2000}.

{\sc shape} is a software package that allows for the quick construction of many structures in 3-D and comparing these results with observed data assists in the interpretation of the object of interest. The geometry of each structure can be easily customised by means of a 3-D module and can be done either analytically or interactively. The application of a velocity field and emission properties to such a 3-D structure model can allow for the generation of 2-D rendered images and P-V arrays, channel maps, light curves, and 1-D spectral line profiles \citep{steffen2011}. Planetary nebulae such as NGC~2392 \citep{garcia2012}, IC~418 \citep{ramos2012b}, HaTr~4 \citep{tyndall2012}, NGC~7026 \citep{clark2013} and Kn~26 \citep{guerrero2013} are all interesting examples that have recently been modelled using {\sc shape}.

A morpho-kinematic model of MyCn~18 was developed to allow for the reconstruction
of its nebular morphology. Our aim, using {\sc shape}, involved creating a 3-D shape resembling the nebula using observed P-V arrays (optical long-slit spectra), which are compared and matched up with synthetic P-V arrays generated from our model. 

Development of the model involved using three main structures; a cylinder (for the outer hourglass), a sphere (for the inner hourglass), and a torus (central `eye' region). A cylinder, with a line type and a segment value of 30 (same in each structure), was used for the outer hourglass shell. A squeeze geometry modifier was then applied to the cylinder to give an hourglass shape with a tight waist of 1.8\arcsec. A velocity modifier, which is a velocity field as a function of position, was also applied to help determine the kinematics of the outer hourglass shell. The vector field of this velocity modifier was set to give a radial Hubble expansion of 0~km~s$^{-1}$ at the centre to 70~km~s$^{-1}$ at the poles using a radial power law. The velocity function is given by $|$\,f\,$|$\,(r) = A + B(r/r$_{0}$) + C(r/r$_{0})^{D}$. The \textit{r} (initial arbitrary radius assigned to the main hourglass structure) parameter is the only constant with a value of 100, whereas all other parameters are variables. The r$_{0}$ parameter is used to give the rate of change of velocity with respect to the radius \textit{r} and has a value of 15. Parameters A and B are 0, whereas C and D are 2.2 and 1.82 respectively. Density modifiers, which set the spatial distribution of density, were used to give the arc-like etchings/rings on the outer hourglass shell. An arbitrary density scaling factor was used where the various structures were modified. A scaling factor of 1.0 was chosen for the main outer hourglass structure. The etchings/arcs, on each lobe, varied in density from 2.0 to 5.0. Two separate density modifiers were applied to the nebular waist (torus structure representing both ring 1 and ring 2). The first of these modifiers varied from 10 to 15 while the other (darkest region of the eye) varied from 15 to 20. All density values were chosen empirically by reference to the HST imagery of MyCn~18 from \citet{Sahai1999}. Other modifiers such as shell thickness, size, and custom vertex were used to complete the outer hourglass structure. It should also be noted that the squeeze and density modifiers were set interactively, whereas the rest were set analytically.  
The inner hourglass was developed using a sphere. Again, a squeeze modifier was applied to give its hourglass shape, and a velocity modifier was set to give a radial expansion from 0~km~s$^{-1}$ at the centre to 30~km~s$^{-1}$ at the poles. The only different modifiers used for this structure were a translation modifier to allow for the offset (0.3\arcsec) of the structure and a rotation modifier to give the slight rotation (5$^{\circ}$) of the inner hourglass with respect to the outer hourglass. The central eye region (nebular waist) of the nebula was constructed using a torus. The only modifiers not used for this structure were the squeeze, translation, and rotation modifiers. The expansion velocity is radially from 0~km~s$^{-1}$ at its centre to 10~km~s$^{-1}$ at its edge. A density modifier was used to give the ring 2 feature, as shown in Fig.~\ref{fig:fig5}a. The value for the inclination of each structure and their features is set to 38$^{\circ}$ on the 2-D rendering module, which corresponds to 90$^{\circ} - 38^{\circ} = 52^{\circ}$. See Table~\ref{table:par} for the parameters used to create the model.

The P-V arrays shown in Fig.~\ref{fig:mes} (AAT \& NTT arrays) and the associated P-V arrays from \citet{o'connor2000} were used as the basis for the reconstruction process. Those observed spectra from Fig. 11 are shown at two different contrasts to display the spectroscopic features of MyCn~18. The stretches for the AAT spectra, from top to centre, are 0--60,000 and 0--15,000 counts respectively, and the stretches for the NTT spectra, from top to centre, are 0--40,000 and 0--10,000 counts, respectively. Also shown in Fig.~\ref{fig:mes} are synthetic P-V arrays, which were generated from the reconstructed model of MyCn~18 by means of a rendering module. The features of the synthetic P-V arrays reveal more detail than those of the observed arrays because of the lesser seeing for the observed data on those given nights. It was for this reason that the P-V arrays from \citet{o'connor2000} needed to be used also. The discrepancies in the observed data were brought out more clearly in the modelled arrays by varying the convolution (seeing) parameters on the rendering module of {\sc shape}. Figs.~\ref{fig:mes} and \ref{fig:shape} were colour-inverted and grey-scaled for clear presentation.

%

\begin{table}
\caption{\label{table:par}The parameters for the best-fit {\sc shape} model of MyCn~18.}
\centering
\begin{tabular}{ll}
\hline\hline\addlinespace[3pt]
Parameter & Value \\ \addlinespace[1pt]
\hline \addlinespace[3pt]
Pole-to-pole length (along its major axis) & 18\arcsec \\ \addlinespace[1pt]
Nebular waist (along its minor axis) & 1.8\arcsec \\ \addlinespace[1pt]
Max lobe width (along its minor axis) & 8.5\arcsec \\ \addlinespace[1pt]
Shell thickness & 0.2\arcsec \\ \addlinespace[1pt]
Inclination (w.r.t. plane of sky) & 52$^{\circ}$ \\ \addlinespace[1pt]
Systemic heliocentric radial velocity, $V_{\rm sys}$ & $-$71~km~s$^{-1}$ \\ \addlinespace[1pt]
Max velocity (outer hourglass) & 70~km~s$^{-1}$ \\ \addlinespace[1pt]
Max velocity (inner hourglass) & 30~km~s$^{-1}$ \\ \addlinespace[1pt]
Convolution (seeing) & 1.37\arcsec \\ \addlinespace[1pt]
\hline
\end{tabular}
\tablefoot{The parameters for the velocity of the inner and outer hourglass varied until the observed and synthetic P-V arrays resembled each other closely. The slit width used in the modelling is the same as that used for the observed data.}
\end{table}

%

A systemic heliocentric velocity, ${V_{\rm sys}}$, of ${-71~\rm km~s^{-1}}$ \citep{o'connor2000} was used as
the central velocity in the synthetic P-V arrays. The brightness of the observed P-V arrays varies from low to high for a long-slit placed directly down the major axis of the nebula 
and across its minor axis. The MES arrays (major-axis arrays) show the lobed features, which are more visible in the higher brightness image. The same features are clearly displayed in the model slit 1 synthetic array with better seeing than the observed MES arrays. It is apparent that both the observed and synthetic arrays follow a similar pattern, which closely indicates that the shape of MyCn~18 is an open ended, bilobed, bipolar nebula.

The rates of expansion (in ${\rm km~s^{-1}}$) from the geometric centre out to the end of each lobe and across the waist were determined from the synthetic arrays. Having applied the appropriate velocity components, using the {\sc shape} velocity modifiers, the velocity scale can be centred at $V_{\rm HEL}=-71~{\rm km~s^{-1}}$, the velocity defined with respect to the nebula's frame of reference in which the Sun is at rest. Two distinctly separate velocity features are present in the model slit-1 and MES arrays with the first around ${0^{\prime\prime}}$ offset from ${V_{\rm HEL} = 0}$ to ${-140~\rm km~s^{-1}}$ and the other from ${V_{\rm HEL} = -30}$ to ${ -120}~\rm km~s^{-1}$ for the model slit-1 array. The MES arrays, however, do not show emission out to these velocities, therefore, the P-V arrays from \citet{o'connor2000} [see Fig.~3] were used to help finalise the synthetic arrays. These velocities obtained closely resemble those found by \citet{o'connor2000} for their observed data.

The finished model, shown in Fig.~\ref{fig:shape}a as a 2-D rendered image, has the same inclination (with the top, southeastern, half of the hourglass tilted towards us at an angle of 52$^{\circ}$ with respect to the plane of the sky) and is closely scaled to that shown in Fig.~\ref{fig:shape}b, which is an HST \nii~image for comparison. It should also be noted that the seeing used for the 2-D rendered image is 0.3 arcsec instead of the 1.37 arcsec used when generating the synthetic P-V arrays, and the structures of the observed P-V arrays are smeared due to the 1.37 arcsec seeing, which makes it difficult to get a precise measurement of the expansion.

\section{Discussion}
\label{discussion}

\subsection{Shaping mechanisms and offsets}

Although there is still much to be understood about the complexity of MyCn~18's asymmetry, we here propose explanations as to why this might be. 
Currently, the binary scenario is the favoured mechanism for the creation of bipolar planetary nebulae \citep{morris1987,han1995,soker2000b,balickfrank2002,demarco2009,jones2010,soker2012}. Companion stars may likely exist within the 
central eye region of MyCn~18, however, since they cannot be observed directly, their luminosity must 
be low in comparison to the neighbouring central star, and any continuum surrounding the central region
would certainly obscure their light.  

In relation to the offset of the central star, one proposed theory involves a variable accretion of matter onto the central white dwarf, resulting in a dwarf nova explosion that could have kicked the star off in one direction while producing the disrupted region of the eye in the opposite direction (see N-E region of ring 1 in Fig.~\ref{fig:fig5}b). Assuming momentum conservation of the ejecta$+$star system before and after the explosion, we determine an energy of 7.2 x 10$^{43}$ ergs for a velocity of 600~km s$^{-1}$ \citep{Bryce1997,o'connor2000} and a mass of 10$^{-5} \rm~M_{\odot}$ \citep{Sahai1999} for the knotty ejecta. From Fig.~\ref{fig:geo}, we see that the star is offset from ring 1 by 0.2\arcsec, which corresponds to 640~AU for a distance of 3.2~kpc, and the time taken for the star to move from the nebula's geometric centre to its current position is 1050 yrs (the difference between the age of the nebula and the knotty ejecta: 2700$-$1650 yrs). Using these values, we find that the star would have to be moving at a velocity of $\sim$3 km~s$^{-1}$ over that period. For a 0.6~M$_{\odot}$ white dwarf and a velocity of 3 km~s$^{-1}$, we find an energy of 5.4 x 10$^{43}$ ergs, which closely resembles that of the energy for the ejecta plus star system. 

The nebula's other prominent features are also offset with respect to the geometric centre. The offset, and even the creation, of the inner hourglass may be the result of the formation of the knots since its axis of symmetry appears to follow the path taken by the knots. Our calculation for the kinematical age of the inner hourglass corresponds closely to that of the knots, which further suggests a possible connection between both components.

The younger, high-speed knots might explain the truncated, open-ended
lobes of the outer hourglass. At an earlier epoch, before the creation of the knots, the ends 
of the lobes may have been closed. Some time after, the high-speed ejection of the knots would 
have penetrated the lobed ends, giving rise to their truncation, thus suggesting the younger age
of the knots. This penetration would have occurred when the outer hourglass was smaller than it is now, which could explain for the large opening angle of each lobe \citep[see][]{sahai1998}.

\citet{soker2012} propose that the formation of bipolar lobes of some PNe and pre-PNe are the result of one or more intermediate-luminosity optical transient (ILOT) events. Jets created by the accretion of matter, from an AGB (or extreme-AGB) star, onto a main-sequence companion are launched by an accretion disk, which in turn shapes the bipolar lobes. For MyCn~18 to form from an ILOT event, it would need to have a linear velocity-distance relation, a bipolar structure, an expansion velocity (for the knotty outflow) of $\sim$100--1000~\rm km~s$^{-1}$, and a total kinetic energy of $\sim$10$^{46}$--10$^{49}$~ergs, as noted by \citet{soker2012}. They show that MyCn~18 has all the characteristic properties of an ILOT event apart from the total kinetic energy, which is of the order $\sim$10$^{44}$ ergs, and their energy-time diagram (ETD) shows MyCn~18 to be just outside the optical transient stripe (OTS), making it an exception. MyCn~18 does not appear energetic enough to be a ILOT event. They support the \citet{o'connor2000} view that MyCn~18 was formed by a nova like explosion instead.
However, in the light of the recent frequent discovery of hot Jupiter-type planets, it is possible to generate an ILOT event of significantly lower energy than envisaged by \citet{soker2012}. A planet engulfed at the AGB phase, surviving to spiral through the tenuous envelope to the dense core of the star and finally forming an accretion disk there \citep{nordhaus2006}, gives a natural mechanism for breaking the spherical symmetry of the AGB phase. Subsequently a nova-type explosion event on the degenerate core could have led to the generation of the hypersonic knots and the deviations from cylindrical symmetry of the nebula and the offset location of its central star.

A planet destruction event causing an asymmetric nova explosion could explain features as follows: on the AGB phase, a planet is engulfed and survives long enough to disrupt the symmetry of the nebula and cause its bipolar shape. As the planet spirals to the degenerate core an accretion disk forms and we speculate that the final accretion event of the planet onto the star generates an explosion that is localised on the stellar core. This asymmetric explosion gives a velocity kick to the star, generates the knotty outflow perpendicular to the disk, and causes the disruption to ring 1 in the opposite (NE) direction to the star's movement \cite[this is observed better in the composite colour image from][]{Sahai1999}.

Interestingly, despite the expectation that planets should be readily detectable around white dwarfs \citep{burleigh2001} they have not been observed yet \citep{mustill2012}, which may suggest that many hot Jupiter exoplanets are readily engulfed and destroyed at the AGB phase.
Since there are not enough close binaries to shape the large fraction of non-spherical PNe, this strongly suggests some role for planets in their shaping \citep{demarco2011}.

\subsection{Morpho-kinematic modelling}

Our long-slit spectra provided us with sufficient information about the 3-D structure of the nebula, which allowed us to generate a self-consistent 3-D model. Density distributions and velocity parameters were applied to the individual sub-systems in order to generate the synthetic P-V arrays needed for describing the kinematics of the nebula. This technique, along with the long-slit spectra, enabled us to reconstruct the shape of the nebula. The resultant 3-D {\sc shape} model is in good agreement with that of \citet{Sahai1999,dayal2000}, specifically for the structure and kinematics of this gaseous nebula.

It has been shown, by use of the observed and synthetic P-V arrays, that the expansion rate of the nebula follows that of a Hubble-type expansion, i.e., the velocity of the ionised gas increases linearly with distance from the centre \citep{solf1985,icke1989}. Our values for the expansion rates correspond closely to those found by \citet{o'connor2000}. The kinematical age of MyCn~18 was then found to be $\sim$2700 yrs using these velocities. Although a distance of 3.2~kpc was used to help determine an age for MyCn~18, it is still not accepted
as a confirmed distance, therefore, other techniques are required to narrow down the possible range.

An unambiguous distance measurement requires a proper motion study 
of the high velocity knots of MyCn~18 or hourglass edges. 
The proper motion for a given feature is
$0.02~V_{100}/D_{\rm kpc} ~{\rm arcsec~yr^{-1}}$, where $V_{100}$ is the
velocity in units of $100~{\rm km~s^{-1}}$ and $D_{\rm kpc}$ is the
distance in units of kpc. The original HST images of MyCn~18 were taken
in the mid 1990s so displacement of the $500~{\rm km~s^{-1}}$ knots
should be detectable in a second epoch image even if the distance to
MyCn~18 is up to 3.2~kpc. 
The second epoch which included the NTT EFOSC frames
were analysed to determine the relative positions of the knots. Unfortunately, the exposure times
were much too short to reveal any knots that might have moved. 

\section{Conclusions}
\label{conclusions}

We have carried out an optical and infrared study of MyCn~18 in order to understand the whole structure and dynamics of the nebula from the offset central star to the striking nebula morphology and the distant hypersonic knotty outflow. VLT IR observations of MyCn~18 were combined with earlier optical work, morpho-kinematic modelling, and a search for a close stellar binary companion. We consider that the conventional planetary nebula formation process has been significantly affected by an explosive event that has led to deviations from axial symmetry such as a displacement of the central star and the misalignment of the main nebular structures. We speculate that a hot Jupiter exoplanet was engulfed at the AGB phase of the star's evolution and that the accretion of this material onto the core of the star led to a nova-type explosion that was ultimately responsible for the hypersonic knotty outflow. Given the common nature of hot Jupiters, it can even be suggested that this and other extreme bipolar morphology planetary nebulae are ultimately shaped by one or more exoplanets.

%
 
\begin{acknowledgements}
We gratefully acknowledge the support of the College of Science, NUI Galway (under their PhD fellowship scheme).
Based on observations collected at the European Southern Observatory, Chile,
and proposal numbers are 071.D-0698 and 073.D-0359.
Based on observations made with the NASA/ESA Hubble Space Telescope, obtained from the data archive at the Space Telescope Institute. STScI is operated by the association of Universities for Research in Astronomy, Inc. under the NASA contract NAS 5-26555. This paper makes use of data obtained from the Isaac Newton Group Archive which is maintained as part of the CASU Astronomical Data Centre at the Institute of Astronomy, Cambridge. We thank Eamonn Harvey for his assistance and guidance with the use of {\sc shape}. We also thank the anonymous referee for their thorough review and highly appreciate their comments and suggestions, which significantly contributed to improving the quality of the publication.
\end{acknowledgements}

%
%

\bibliographystyle{aa}
\bibliography{reference_database}

\begin{thebibliography}{82}
\expandafter\ifx\csname natexlab\endcsname\relax\def\natexlab#1{#1}\fi

\bibitem[{{Aleman} \& {Gruenwald}(2004)}]{aleman2004}
{Aleman}, I. \& {Gruenwald}, R. 2004, \apj, 607, 865

\bibitem[{{Bains} {et~al.}(2002){Bains}, {Bryce}, {Calabretta}, \&
  {Stirling}}]{Bains2002}
{Bains}, I., {Bryce}, M., {Calabretta}, M., \& {Stirling}, A.~M. 2002, \mnras,
  337, 401

\bibitem[{{Balick}(1987)}]{balick1987}
{Balick}, B. 1987, \aj, 94, 671

\bibitem[{{Balick} \& {Frank}(2002)}]{balickfrank2002}
{Balick}, B. \& {Frank}, A. 2002, \araa, 40, 439

\bibitem[{{Bernard-Salas} \& {Tielens}(2005)}]{Bernard-Salas2005}
{Bernard-Salas}, J. \& {Tielens}, A.~G.~G.~M. 2005, \aap, 431, 523

\bibitem[{{Bryce} {et~al.}(1997){Bryce}, {Lopez}, {Holloway}, \&
  {Meaburn}}]{Bryce1997}
{Bryce}, M., {Lopez}, J.~A., {Holloway}, A.~J., \& {Meaburn}, J. 1997, \apjl,
  487, L161+

\bibitem[{{Burleigh} {et~al.}(2001){Burleigh}, {Clarke}, \&
  {Hodgkin}}]{burleigh2001}
{Burleigh}, M., {Clarke}, F., \& {Hodgkin}, S. 2001, in Techniques for the
  Detection of Planets and Life beyond the Solar System, ed. W.~R.~F. {Dent},
  15

\bibitem[{{Burton} {et~al.}(1992){Burton}, {Hollenbach}, \&
  {Tielens}}]{Burton1992}
{Burton}, M.~G., {Hollenbach}, D.~J., \& {Tielens}, A.~G.~G. 1992, \apj, 399,
  563

\bibitem[{{Cavanagh} {et~al.}(2008){Cavanagh}, {Jenness}, {Economou}, \&
  {Currie}}]{Cavanagh2008}
{Cavanagh}, B., {Jenness}, T., {Economou}, F., \& {Currie}, M.~J. 2008,
  Astronomische Nachrichten, 329, 295

\bibitem[{{Chevalier} \& {Luo}(1994)}]{chevalier1994}
{Chevalier}, R.~A. \& {Luo}, D. 1994, \apj, 421, 225

\bibitem[{{Clark} {et~al.}(2013){Clark}, {L{\'o}pez}, {Steffen}, \&
  {Richer}}]{clark2013}
{Clark}, D.~M., {L{\'o}pez}, J.~A., {Steffen}, W., \& {Richer}, M.~G. 2013,
  \aj, 145, 57

\bibitem[{{Corradi} \& {Schwarz}(1993)}]{corradi1993}
{Corradi}, R.~L.~M. \& {Schwarz}, H.~E. 1993, \aap, 268, 714

\bibitem[{{Cox} {et~al.}(1998){Cox}, {Boulanger}, {Huggins}, {Tielens},
  {Forveille}, {Bachiller}, {Cesarsky}, {Jones}, {Young}, {Roelfsema}, \&
  {Cernicharo}}]{Cox1998}
{Cox}, P., {Boulanger}, F., {Huggins}, P.~J., {et~al.} 1998, \apjl, 495, L23+

\bibitem[{{Currie} \& {Berry}(2000)}]{currie2000}
{Currie}, M.~J. \& {Berry}, D.~S. 2000, Starlink User Note 95

\bibitem[{{Currie} \& {Cavanagh}(2004)}]{Currie2004}
{Currie}, M.~J. \& {Cavanagh}, B. 2004, ORAC-DR - imaging data reduction,
  starlink User Note 232.8

\bibitem[{{Davis} {et~al.}(2003){Davis}, {Smith}, {Stern}, {Kerr}, \&
  {Chiar}}]{Davis2003}
{Davis}, C.~J., {Smith}, M.~D., {Stern}, L., {Kerr}, T.~H., \& {Chiar}, J.~E.
  2003, \mnras, 344, 262

\bibitem[{{Dayal} {et~al.}(1997){Dayal}, {Sahai}, {Trauger}, {Hora}, {Fazio},
  {Hoffmann}, {Bieging}, {Deutsch}, \& {Latter}}]{dayal1997}
{Dayal}, A., {Sahai}, R., {Trauger}, J., {et~al.} 1997, in Bulletin of the
  American Astronomical Society, Vol.~29, American Astronomical Society Meeting
  Abstracts, 1234

\bibitem[{{Dayal} {et~al.}(2000){Dayal}, {Sahai}, {Watson}, {Trauger},
  {Burrows}, {Stapelfeldt}, \& {Gallagher}}]{dayal2000}
{Dayal}, A., {Sahai}, R., {Watson}, A.~M., {et~al.} 2000, \aj, 119, 315

\bibitem[{{De Marco}(2009)}]{demarco2009}
{De Marco}, O. 2009, \pasp, 121, 316

\bibitem[{{De Marco} \& {Soker}(2011)}]{demarco2011}
{De Marco}, O. \& {Soker}, N. 2011, \pasp, 123, 402

\bibitem[{{Dekker} {et~al.}(1986){Dekker}, {Delabre}, \&
  {Dodorico}}]{dekker1986}
{Dekker}, H., {Delabre}, B., \& {Dodorico}, S. 1986, in Society of
  Photo-Optical Instrumentation Engineers (SPIE) Conference Series, Vol. 627,
  Society of Photo-Optical Instrumentation Engineers (SPIE) Conference Series,
  ed. D.~L. {Crawford}, 339--348

\bibitem[{{Disney} \& {Wallace}(1982)}]{disney1982}
{Disney}, M.~J. \& {Wallace}, P.~T. 1982, \qjras, 23, 485

\bibitem[{{Douchin} {et~al.}(2012){Douchin}, {De}, {Jacoby}, {Hillwig}, {Frew},
  {Bojicic}, {Jasniewicz}, \& {Parker}}]{douchin2012}
{Douchin}, D., {De}, M.~O., {Jacoby}, G.~H., {et~al.} 2012, in SF2A-2012:
  Proceedings of the Annual meeting of the French Society of Astronomy and
  Astrophysics, ed. S.~{Boissier}, P.~{de Laverny}, N.~{Nardetto}, R.~{Samadi},
  D.~{Valls-Gabaud}, \& H.~{Wozniak}, 325--327

\bibitem[{{Eyermann} \& {Speck}(2004)}]{eyermann2004}
{Eyermann}, S.~E. \& {Speck}, A.~K. 2004, in Bulletin of the American
  Astronomical Society, Vol.~36, American Astronomical Society Meeting
  Abstracts, 138.11

\bibitem[{{Garc{\'{\i}}a-D{\'{\i}}az}
  {et~al.}(2012){Garc{\'{\i}}a-D{\'{\i}}az}, {L{\'o}pez}, {Steffen}, \&
  {Richer}}]{garcia2012}
{Garc{\'{\i}}a-D{\'{\i}}az}, M.~T., {L{\'o}pez}, J.~A., {Steffen}, W., \&
  {Richer}, M.~G. 2012, \apj, 761, 172

\bibitem[{{Garc{\'{\i}}a-Segura} {et~al.}(1999){Garc{\'{\i}}a-Segura},
  {Langer}, {R{\'o}{\.z}yczka}, \& {Franco}}]{garcia1999}
{Garc{\'{\i}}a-Segura}, G., {Langer}, N., {R{\'o}{\.z}yczka}, M., \& {Franco},
  J. 1999, \apj, 517, 767

\bibitem[{{Guerrero} {et~al.}(2013){Guerrero}, {Miranda}, {Ramos-Larios}, \&
  {V{\'a}zquez}}]{guerrero2013}
{Guerrero}, M.~A., {Miranda}, L.~F., {Ramos-Larios}, G., \& {V{\'a}zquez}, R.
  2013, \aap, 551, A53

\bibitem[{{Hadrava}(2006)}]{hadrava2006}
{Hadrava}, P. 2006, \aap, 448, 1149

\bibitem[{{Han} {et~al.}(1995){Han}, {Podsiadlowski}, \& {Eggleton}}]{han1995}
{Han}, Z., {Podsiadlowski}, P., \& {Eggleton}, P.~P. 1995, \mnras, 272, 800

\bibitem[{{Henney} {et~al.}(2007){Henney}, {Williams}, {Ferland}, {Shaw}, \&
  {O'Dell}}]{Henney2007}
{Henney}, W.~J., {Williams}, R.~J.~R., {Ferland}, G.~J., {Shaw}, G., \&
  {O'Dell}, C.~R. 2007, \apjl, 671, L137

\bibitem[{{Hora} \& {Latter}(1996)}]{hora1996}
{Hora}, J.~L. \& {Latter}, W.~B. 1996, in Bulletin of the American Astronomical
  Society, Vol.~28, American Astronomical Society Meeting Abstracts, 1402

\bibitem[{{Hora} {et~al.}(2005){Hora}, {Latter}, {Marengo}, {Fazio}, {Allen},
  \& {Pipher}}]{hora2005}
{Hora}, J.~L., {Latter}, W.~B., {Marengo}, M., {et~al.} 2005, in Bulletin of
  the American Astronomical Society, Vol.~37, American Astronomical Society
  Meeting Abstracts \#206, 493

\bibitem[{{Icke} {et~al.}(1989){Icke}, {Preston}, \& {Balick}}]{icke1989}
{Icke}, V., {Preston}, H.~L., \& {Balick}, B. 1989, \aj, 97, 462

\bibitem[{{Jones} {et~al.}(2010){Jones}, {Lloyd}, {Santander-Garc{\'{\i}}a},
  {L{\'o}pez}, {Meaburn}, {Mitchell}, {O'Brien}, {Pollacco},
  {Rubio-D{\'{\i}}ez}, \& {Vaytet}}]{jones2010}
{Jones}, D., {Lloyd}, M., {Santander-Garc{\'{\i}}a}, M., {et~al.} 2010, \mnras,
  408, 2312

\bibitem[{{Jones} {et~al.}(2012){Jones}, {Mitchell}, {Lloyd}, {Pollacco},
  {O'Brien}, {Meaburn}, \& {Vaytet}}]{jones2012}
{Jones}, D., {Mitchell}, D.~L., {Lloyd}, M., {et~al.} 2012, \mnras, 420, 2271

\bibitem[{{Jordan} {et~al.}(2012){Jordan}, {Bagnulo}, {Werner}, \&
  {O'Toole}}]{jordan2012}
{Jordan}, S., {Bagnulo}, S., {Werner}, K., \& {O'Toole}, S.~J. 2012, \aap, 542,
  A64

\bibitem[{{Kastner} {et~al.}(2000){Kastner}, {Gatley}, \&
  {Weintraub}}]{kastner2000}
{Kastner}, J.~H., {Gatley}, I., \& {Weintraub}, D.~A. 2000, in Astronomical
  Society of the Pacific Conference Series, Vol. 199, Asymmetrical Planetary
  Nebulae II: From Origins to Microstructures, ed. J.~H. {Kastner}, N.~{Soker},
  \& S.~{Rappaport}, 355

\bibitem[{{Kastner} {et~al.}(1998){Kastner}, {Henn}, {Weintraub}, {Gatley}, \&
  {Siebenmorgen}}]{kastner1998}
{Kastner}, J.~H., {Henn}, L., {Weintraub}, D.~A., {Gatley}, I., \&
  {Siebenmorgen}, R. 1998, in Bulletin of the American Astronomical Society,
  Vol.~30, American Astronomical Society Meeting Abstracts, 1274

\bibitem[{{Koller} \& {Kimeswenger}(2000)}]{koller2000}
{Koller}, J. \& {Kimeswenger}, S. 2000, Thermal Emission Spectroscopy and
  Analysis of Dust, Disks, and Regoliths, 196, 23

\bibitem[{{Kwok}(1982)}]{kwok1982}
{Kwok}, S. 1982, \apj, 258, 280

\bibitem[{{Kwok} {et~al.}(1978){Kwok}, {Purton}, \& {Fitzgerald}}]{kwok1978}
{Kwok}, S., {Purton}, C.~R., \& {Fitzgerald}, P.~M. 1978, \apjl, 219, L125

\bibitem[{{Lee} {et~al.}(2007){Lee}, {Stanghellini}, {Ferrario}, \&
  {Wickramasinghe}}]{Lee2007}
{Lee}, T.-H., {Stanghellini}, L., {Ferrario}, L., \& {Wickramasinghe}, D.~T.
  2007, in Astronomical Society of the Pacific Conference Series, Vol. 372,
  15th European Workshop on White Dwarfs, ed. {R.~Napiwotzki \&
  M.~R.~Burleigh}, 173

\bibitem[{{Lenzen} {et~al.}(2003){Lenzen}, {Hartung}, {Brandner}, {Finger},
  {Hubin}, {Lacombe}, {Lagrange}, {Lehnert}, {Moorwood}, \&
  {Mouillet}}]{Lenzen2003}
{Lenzen}, R., {Hartung}, M., {Brandner}, W., {et~al.} 2003, in Society of
  Photo-Optical Instrumentation Engineers (SPIE) Conference Series, Vol. 4841,
  Society of Photo-Optical Instrumentation Engineers (SPIE) Conference Series,
  ed. M.~{Iye} \& A.~F.~M. {Moorwood}, 944--952

\bibitem[{{Likkel} {et~al.}(2000){Likkel}, {Kindt}, {Dinerstein}, \&
  {Lester}}]{likkel2000}
{Likkel}, L., {Kindt}, A., {Dinerstein}, H.~L., \& {Lester}, D.~F. 2000, in
  Astronomical Society of the Pacific Conference Series, Vol. 199, Asymmetrical
  Planetary Nebulae II: From Origins to Microstructures, ed. J.~H. {Kastner},
  N.~{Soker}, \& S.~{Rappaport}, 333

\bibitem[{{Maihara} {et~al.}(1993){Maihara}, {Iwamuro}, {Yamashita}, {Hall},
  {Cowie}, {Tokunaga}, \& {Pickles}}]{maihara1993}
{Maihara}, T., {Iwamuro}, F., {Yamashita}, T., {et~al.} 1993, \pasp, 105, 940

\bibitem[{{Matsuura} {et~al.}(2007){Matsuura}, {Speck}, {Smith}, {Zijlstra},
  {Viti}, {Lowe}, {Redman}, {Wareing}, \& {Lagadec}}]{Matsuura2007}
{Matsuura}, M., {Speck}, A.~K., {Smith}, M.~D., {et~al.} 2007, \mnras, 382,
  1447

\bibitem[{{Matsuura} \& {Zijlstra}(2005)}]{matsuura2005}
{Matsuura}, M. \& {Zijlstra}, A. 2005, in High Resolution Infrared Spectroscopy
  in Astronomy, ed. H.~U. {K{\"a}ufl}, R.~{Siebenmorgen}, \& A.~{Moorwood},
  423--426

\bibitem[{{Meaburn} {et~al.}(1984){Meaburn}, {Blundell}, {Carling}, {Gregory},
  {Keir}, \& {Wynne}}]{meaburn1984}
{Meaburn}, J., {Blundell}, B., {Carling}, R., {et~al.} 1984, \mnras, 210, 463

\bibitem[{{Moorwood} {et~al.}(1999){Moorwood}, {Cuby}, {Ballester},
  {Biereichel}, {Brynnel}, {Conzelmann}, {Delabre}, {Devillard}, {van
  Dijsseldonk}, {Finger}, {Gemperlein}, {Lidman}, {Herlin}, {Huster},
  {Knudstrup}, {Lizon}, {Mehrgan}, {Meyer}, {Nicolini}, {Silber}, {Spyromilio},
  \& {Stegmeier}}]{Moorwood1999}
{Moorwood}, A., {Cuby}, J.-G., {Ballester}, P., {et~al.} 1999, The Messenger,
  95, 1

\bibitem[{{Morris}(1987)}]{morris1987}
{Morris}, M. 1987, \pasp, 99, 1115

\bibitem[{{Mustill} \& {Villaver}(2012)}]{mustill2012}
{Mustill}, A.~J. \& {Villaver}, E. 2012, \apj, 761, 121

\bibitem[{{Nordhaus} \& {Blackman}(2006)}]{nordhaus2006}
{Nordhaus}, J. \& {Blackman}, E.~G. 2006, \mnras, 370, 2004

\bibitem[{{O'Connor} {et~al.}(2000){O'Connor}, {Redman}, {Holloway}, {Bryce},
  {L{\'o}pez}, \& {Meaburn}}]{o'connor2000}
{O'Connor}, J.~A., {Redman}, M.~P., {Holloway}, A.~J., {et~al.} 2000, \apj,
  531, 336

\bibitem[{{Oliva} \& {Origlia}(1992)}]{oliva1992}
{Oliva}, E. \& {Origlia}, L. 1992, \aap, 254, 466

\bibitem[{{Pascoli} \& {Lahoche}(2008)}]{pascoli2008}
{Pascoli}, G. \& {Lahoche}, L. 2008, \pasp, 120, 1267

\bibitem[{{Pickles}(1998)}]{Pickles1998}
{Pickles}, A.~J. 1998, \pasp, 110, 863

\bibitem[{{Ramos-Larios} {et~al.}(2008){Ramos-Larios}, {Guerrero}, \&
  {Miranda}}]{ramos2008}
{Ramos-Larios}, G., {Guerrero}, M.~A., \& {Miranda}, L.~F. 2008, \aj, 135, 1441

\bibitem[{{Ramos-Larios} {et~al.}(2012{\natexlab{a}}){Ramos-Larios},
  {Guerrero}, {Su{\'a}rez}, {Miranda}, \& {G{\'o}mez}}]{ramos2012}
{Ramos-Larios}, G., {Guerrero}, M.~A., {Su{\'a}rez}, O., {Miranda}, L.~F., \&
  {G{\'o}mez}, J.~F. 2012{\natexlab{a}}, \aap, 545, A20

\bibitem[{{Ramos-Larios} {et~al.}(2012{\natexlab{b}}){Ramos-Larios},
  {V{\'a}zquez}, {Guerrero}, {Olgu{\'{\i}}n}, {Marquez-Lugo}, \&
  {Bravo-Alfaro}}]{ramos2012b}
{Ramos-Larios}, G., {V{\'a}zquez}, R., {Guerrero}, M.~A., {et~al.}
  2012{\natexlab{b}}, \mnras, 423, 3753

\bibitem[{{Rousselot} {et~al.}(2000){Rousselot}, {Lidman}, {Cuby}, {Moreels},
  \& {Monnet}}]{Rousselot2000}
{Rousselot}, P., {Lidman}, C., {Cuby}, J.-G., {Moreels}, G., \& {Monnet}, G.
  2000, \aap, 354, 1134

\bibitem[{{Rousset} {et~al.}(2003){Rousset}, {Lacombe}, {Puget}, {Hubin},
  {Gendron}, {Fusco}, {Arsenault}, {Charton}, {Feautrier}, {Gigan}, {Kern},
  {Lagrange}, {Madec}, {Mouillet}, {Rabaud}, {Rabou}, {Stadler}, \&
  {Zins}}]{Rousset2003}
{Rousset}, G., {Lacombe}, F., {Puget}, P., {et~al.} 2003, in Society of
  Photo-Optical Instrumentation Engineers (SPIE) Conference Series, Vol. 4839,
  Society of Photo-Optical Instrumentation Engineers (SPIE) Conference Series,
  ed. P.~L. {Wizinowich} \& D.~{Bonaccini}, 140--149

\bibitem[{{Sahai} {et~al.}(1999){Sahai}, {Dayal}, {Watson}, {Trauger},
  {Stapelfeldt}, {Burrows}, {Gallagher}, {Scowen}, {Hester}, {Evans},
  {Ballester}, {Clarke}, {Crisp}, {Griffiths}, {Hoessel}, {Holtzman}, {Krist},
  \& {Mould}}]{Sahai1999}
{Sahai}, R., {Dayal}, A., {Watson}, A.~M., {et~al.} 1999, \aj, 118, 468

\bibitem[{{Sahai} {et~al.}(2009){Sahai}, {Sugerman}, \& {Hinkle}}]{sahai2009}
{Sahai}, R., {Sugerman}, B.~E.~K., \& {Hinkle}, K. 2009, \apj, 699, 1015

\bibitem[{{Sahai} \& {Trauger}(1998)}]{sahai1998}
{Sahai}, R. \& {Trauger}, J.~T. 1998, \aj, 116, 1357

\bibitem[{{Schwarz} {et~al.}(1992){Schwarz}, {Corradi}, \&
  {Melnick}}]{schwarz1992}
{Schwarz}, H.~E., {Corradi}, R.~L.~M., \& {Melnick}, J. 1992, \aaps, 96, 23

\bibitem[{{Sellgren} {et~al.}(2008){Sellgren}, {Bromm}, \&
  {Dinerstein}}]{sellgren2008}
{Sellgren}, K., {Bromm}, V., \& {Dinerstein}, H. 2008, Spitzer Proposal, 50179

\bibitem[{{Soker} \& {Harpaz}(1992)}]{soker1992}
{Soker}, N. \& {Harpaz}, A. 1992, \pasp, 104, 923

\bibitem[{{Soker} \& {Kashi}(2012)}]{soker2012}
{Soker}, N. \& {Kashi}, A. 2012, \apj, 746, 100

\bibitem[{{Soker} \& {Rappaport}(2000)}]{soker2000b}
{Soker}, N. \& {Rappaport}, S. 2000, \apj, 538, 241

\bibitem[{{Soker} \& {Rappaport}(2001)}]{soker2001}
{Soker}, N. \& {Rappaport}, S. 2001, \apj, 557, 256

\bibitem[{{Solf} \& {Ulrich}(1985)}]{solf1985}
{Solf}, J. \& {Ulrich}, H. 1985, \aap, 148, 274

\bibitem[{{Stanghellini} {et~al.}(2008){Stanghellini}, {Shaw}, \&
  {Villaver}}]{stanghellini2008}
{Stanghellini}, L., {Shaw}, R.~A., \& {Villaver}, E. 2008, \apj, 689, 194

\bibitem[{Steffen {et~al.}(2011)Steffen, Koning, Wenger, Morisset, \&
  Magnor}]{steffen2011}
Steffen, W., Koning, N., Wenger, S., Morisset, C., \& Magnor, M. 2011, IEEE
  Transactions on Visualization and Computer Graphics, 17, 454

\bibitem[{{Tenenbaum} {et~al.}(2009){Tenenbaum}, {Milam}, {Woolf}, \&
  {Ziurys}}]{tenenbaum2009}
{Tenenbaum}, E.~D., {Milam}, S.~N., {Woolf}, N.~J., \& {Ziurys}, L.~M. 2009,
  \apjl, 704, L108

\bibitem[{{Trammell} {et~al.}(1995){Trammell}, {Dinerstein}, \&
  {Goodrich}}]{trammell1995}
{Trammell}, S.~R., {Dinerstein}, H.~L., \& {Goodrich}, R.~W. 1995, in Annals of
  the Israel Physical Society, Vol.~11, Asymmetrical Planetary Nebulae, ed.
  A.~{Harpaz} \& N.~{Soker}, 37

\bibitem[{{Tyndall} {et~al.}(2012){Tyndall}, {Jones}, {Lloyd}, {O'Brien}, \&
  {Pollacco}}]{tyndall2012}
{Tyndall}, A.~A., {Jones}, D., {Lloyd}, M., {O'Brien}, T.~J., \& {Pollacco}, D.
  2012, \mnras, 422, 1804

\bibitem[{{Vlemmings}(2012)}]{vlemmings2012}
{Vlemmings}, W. 2012, in IAU Symposium, Vol. 283, IAU Symposium, 176--179

\bibitem[{{Vlemmings}(2011)}]{vlemmings2011}
{Vlemmings}, W.~H.~T. 2011, in Asymmetric Planetary Nebulae 5 Conference

\bibitem[{{Volk}(2003)}]{volk2003}
{Volk}, K. 2003, in IAU Symposium, Vol. 209, Planetary Nebulae: Their Evolution
  and Role in the Universe, ed. S.~{Kwok}, M.~{Dopita}, \& R.~{Sutherland}, 281

\bibitem[{{Wang} {et~al.}(2006){Wang}, {Muthumariappan}, \& {Kwok}}]{wang2006}
{Wang}, M.-Y., {Muthumariappan}, C., \& {Kwok}, S. 2006, in IAU Symposium, Vol.
  234, Planetary Nebulae in our Galaxy and Beyond, ed. M.~J. {Barlow} \& R.~H.
  {M{\'e}ndez}, 537--538

\bibitem[{{Whitelock}(1985)}]{whitelock1985}
{Whitelock}, P.~A. 1985, \mnras, 213, 59

\bibitem[{{Yungelson} {et~al.}(1993){Yungelson}, {Tutukov}, \&
  {Livio}}]{yungelson1993}
{Yungelson}, L.~R., {Tutukov}, A.~V., \& {Livio}, M. 1993, \apj, 418, 794

\end{thebibliography}

\clearpage
\begin{appendix}
\section*{Appendix A: {\sc shape} images of MyCn~18}
Shown above is the {\sc shape} model for MyCn~18 from different viewing angles. The green, yellow, and purple line represent the top view, right view, and axis of symmetry of the nebula, respectively. The top left view shows the nebula with its southern pole tilted towards us at an angle of 52$^{\circ}$ with respect to the plane of the sky. From top to bottom on this view represents a NW to SE direction. The view on the top middle is a close in of the first showing detail of the inner part of the nebula.

%
\begin{figure}
\centering
\includegraphics[width=17.5cm]{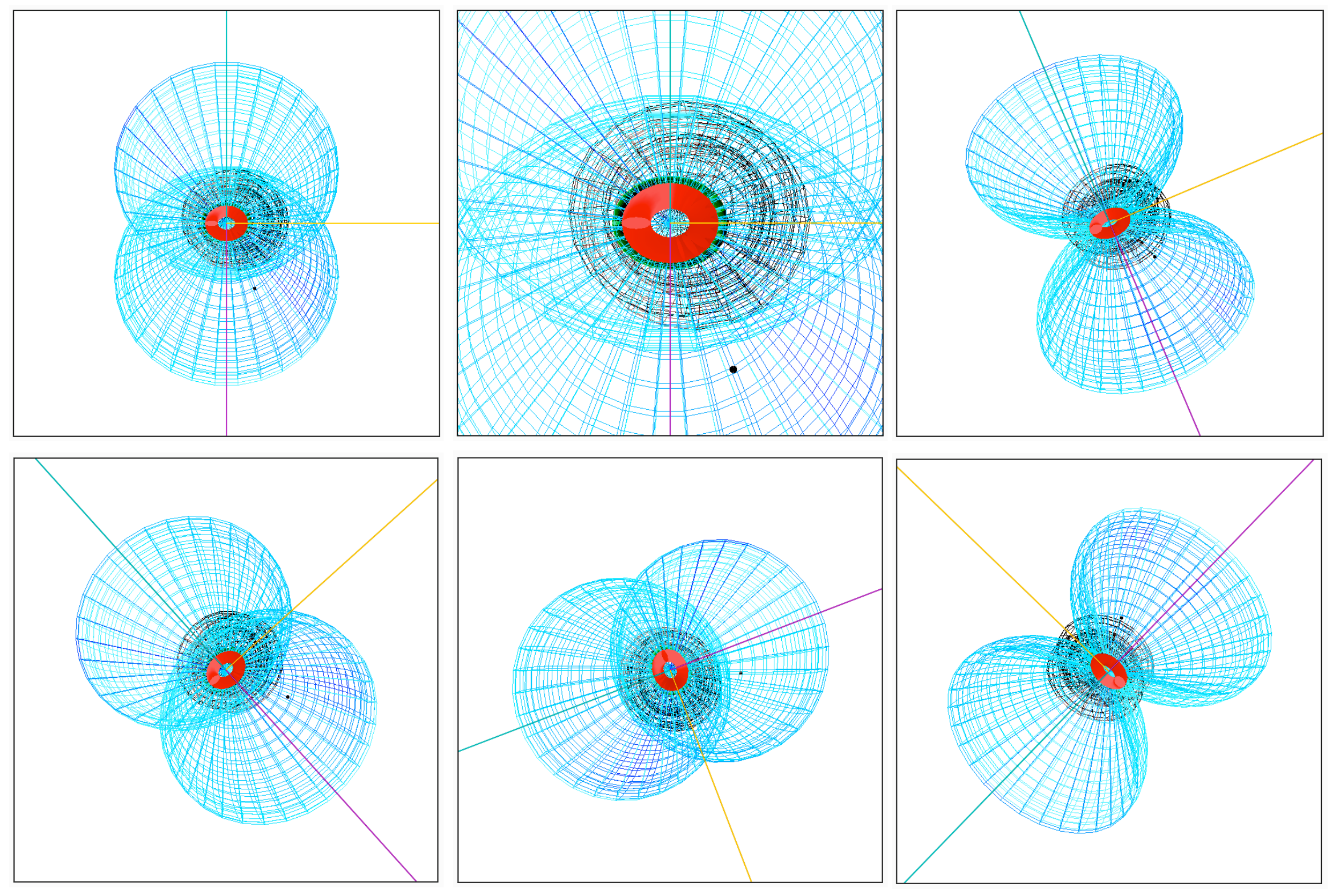}
\caption{{\sc shape} model of MyCn 18 from different points of view.}
\label{fig:mesh}
\end{figure}

%

\end{appendix}

\end{document}